\documentclass[acmsmall]{acmart}
\AtBeginDocument{%
  \providecommand\BibTeX{{%
    \normalfont B\kern-0.5em{\scshape i\kern-0.25em b}\kern-0.8em\TeX}}}

\setcopyright{acmcopyright}
\copyrightyear{2018}
\acmYear{2018}
\acmDOI{XXXXXXX.XXXXXXX}

\acmJournal{JACM}
\acmVolume{37}
\acmNumber{4}
\acmArticle{111}
\acmMonth{8}
\usepackage{multirow}
\usepackage{bbding}
\usepackage{graphicx}
\usepackage{float}
\usepackage{subfigure}
\usepackage{longtable}

\begin{document}

\title{Contrastive Self-supervised Learning in Recommender Systems: A Survey}

\author{Mengyuan Jing}
\email{jingmy@sjtu.edu.cn}
\affiliation{%
  \institution{Shanghai Jiao Tong University}
  \streetaddress{No.800 Dongchuan Road, Minhang District}
  \city{Shanghai}
  \state{Shanghai}
  \country{China}
  \postcode{200240}
}
\author{Yanmin Zhu}\authornote{corresponding author}
\email{yzhu@sjtu.edu.cn}
\affiliation{%
  \institution{Shanghai Jiao Tong University}
  \streetaddress{No.800 Dongchuan Road, Minhang District}
  \city{Shanghai}
  \state{Shanghai}
  \country{China}
  \postcode{200240}
}
\author{Tianzi Zang}
\email{zangtianzi@sjtu.edu.cn}
\affiliation{%
  \institution{Shanghai Jiao Tong University}
  \streetaddress{No.800 Dongchuan Road, Minhang District}
  \city{Shanghai}
  \state{Shanghai}
  \country{China}
  \postcode{200240}
}
\author{Ke Wang}
\email{onecall@sjtu.edu.cn}
\affiliation{%
  \institution{Shanghai Jiao Tong University}
  \streetaddress{No.800 Dongchuan Road, Minhang District}
  \city{Shanghai}
  \state{Shanghai}
  \country{China}
  \postcode{200240}
}

\renewcommand{\shortauthors}{M. Jing et al.}

\begin{abstract}
 Deep learning-based recommender systems have achieved remarkable success in recent years. However, these methods usually heavily rely on labeled data (i.e., user-item interactions), suffering from problems such as data sparsity and cold-start. Self-supervised learning, an emerging paradigm that extracts information from unlabeled data, provides insights into addressing these problems. Specifically, contrastive self-supervised learning, due to its flexibility and promising performance, has attracted considerable interest and recently become a dominant branch in self-supervised learning-based recommendation methods. In this survey, we provide an up-to-date and comprehensive review of current contrastive self-supervised learning-based recommendation methods. Firstly, we propose a unified framework for these methods. We then introduce a taxonomy based on the key components of the framework, including view generation strategy, contrastive task, and contrastive objective. For each component, we provide detailed descriptions and discussions to guide the choice of the appropriate method. Finally, we outline open issues and promising directions for future research.
\end{abstract}



\maketitle

\section{Introduction}

Recommender systems, as the most effective way to alleviate information overloading, have been an indispensable tool in daily life~\cite{10.1145/2959100.2959190, zhang2019deep}.
They are intensively employed in a broad range of online services such as e-commerce platforms, social media, and music platforms. Owing to the ability to effectively capture the user-item relationships, deep learning techniques have been widely used in recommender systems~\cite{cheng2016wide, he2017neural}. Despite their effectiveness, most deep learning-based methods focus on supervised learning settings. The recommendation model is trained with abundant labeled data (i.e., user-item interactions). However, user-item interaction records are very sparse compared to the interaction space~\cite{bayer2017generic, he2016ups}. 
Hence, these methods usually suffer from the problem of data sparsity~\cite{wuSelfsupervisedGraphLearning2021}. Meanwhile, these methods are prone to the problem of over-fitting and generalization error~\cite{liu2021self}.

Self-supervised learning (SSL)~\cite{liu2021self}, as a novel learning paradigm, provides new insights to overcome aforementioned problems. The basic idea of SSL is to acquire transferable knowledge from the data itself without the need for manually annotated labels. This is achieved by solving auxiliary tasks (named pretext tasks). The acquired knowledge is then used in downstream tasks. Due to its efficiency, SSL has been widely used in many fields such as computer vision (CV)~\cite{chen2020simclr, hjelm2018learning, oord2018representation}, natural language processing (NLP)~\cite{gao2021simcse, logeswaran2018efficient} and graph learning~\cite{wu2022discovering, DGI}. Inspired by the success of SSL in other fields, there is growing interest in applying SSL to the area of recommendation.

Existing SSL-based recommendation methods can be classified into generative, contrastive, and predictive methods~\cite{SSLSurvey}. However, generative self-supervised learning is memory-consuming when trained on large-scale datasets. Predictive self-supervised learning often requires domain knowledge to generate labels, leading to increased costs and reduced generalization performance. In contrast, contrastive self-supervised learning (CL for brevity) has lightweight models and flexible designs since it is independent of the encoder structure and typically requires no domain knowledge. As a result, CL-based methods have witnessed significant growth in recent years, emerging as the dominant approach among SSL-based recommendation methods. Furthermore, the number of publications related to CL-based recommendations exceeds 50\% of the total number of SSL-based recommendation publications in the ACM Digital Library\footnote{https://dl.acm.org/}. Considering this increasing trend, we aim to provide a timely and comprehensive review to summarize these CL-based methods in this paper.

Although there have been several reviews~\cite{jaiswal2020survey, khan2022contrastive} on contrastive learning, they mainly focus on methods in CV and NLP without reviewing CL-based recommendation methods. However, due to the uniqueness of the recommendation, it is difficult to apply existing CL-based methods from other fields to recommendation. Specifically, in CV/NLP, models usually deal with dense input data and treat each data instance as isolated. However, in recommender systems, the input data are extremely sparse (e.g., one-hot ID and categorical features of users/items) and there is homophily between users or items. Moreover, various recommendation tasks are unique to recommender systems, such as bundle recommendation and multi-behavior recommendation. 
Furthermore, several reviews on SSL-based graph learning~\cite{graph_survey1, graph_survey2, liu2022graph_survey_3} also include some CL-based recommendation methods. However, these reviews only provide a limited introduction and lack in-depth discussions on recommendation. 
Therefore, these reviews cannot provide sufficient insights into CL-based recommendation. Considering the unique characteristics of recommendation, a comprehensive survey is necessary to thoroughly review CL-based recommendation methods.

In the field of recommendation, the most relevant survey is~\cite{SSLSurvey}. This survey reviews SSL-based recommendation methods, including some CL-based methods. Compared to~\cite{SSLSurvey}, our survey purely focuses on CL-based recommendation and provides a more comprehensive and detailed analysis of this topic. Specifically, our survey has the following differences. Firstly, we present a more rational, fine-grained, and comprehensive taxonomy. For instance, we add model-based augmentation methods and methods without augmentation to view generation strategies and categorize contrastive tasks based on the characteristics of the contrastive instances. Secondly, we provide in-depth analyses of different options for key components of CL-based recommendation methods, guiding the selection of these components. This critical discussion is not present in~\cite{SSLSurvey}. Finally, because of the increasing popularity of CL-based recommendation methods, we provide a more up-to-date review that summarizes recently published studies that were not included in~\cite{SSLSurvey}.

To sum up, the key contributions of this paper are summarized as follows:

\begin{itemize}
  \item We propose a general framework to unify the CL-based methods for recommendation. Based on the framework, we review existing research according to three key components: view generation, pretext task, and contrastive objective. 
  \item We provide an up-to-date and comprehensive review of CL-based recommendation methods. We provide detailed descriptions and discussions for each key component to guide the choice of the appropriate method. We also introduce the relevant background knowledge to help readers easily understand CL-based recommendation. 
  \item We identify the limitations of existing research and propose promising future directions for CL-based recommendation to inspire new research.
\end{itemize}

\textbf{Paper Collection.}
We first adopt Google Scholar as the main search engine to collect related papers.
Then, we search for related work from top-tier conferences and journals, such as SIGIR, KDD, WWW, AAAI, IJCAI, WSDM, CIKM, NuerIPS, ICML, TKDE, TOIS, etc. Specifically, we search with keywords including  "self-supervised", "contrastive" in combination with  "recommend", "collaborative filtering".
To prevent omissions of relevant work, we further look through the references of each paper.

\textbf{Survey Organization.}
The remainder of the survey is organized as follows. In Section~\ref{sec:background}, we introduce background knowledge.
We then introduce the unified framework and taxonomy in Section~\ref{sec:taxonomy}.
Section~\ref{sec:view_gen}, Section~\ref{sec:pretext_task} and Section~\ref{sec:obj} are the main contents, which review contrastive learning in recommender systems. 
In Section~\ref{sec:future}, we discuss the open problems and future directions. Finally, we conclude the survey in Section~\ref{sec:conclusion}.

\section{Background}\label{sec:background}
In this section, we introduce essential background knowledge about CL-based recommendation.
First, we provide the definitions of relevant concepts.
Then, we give a brief introduction to contrastive learning.
At last, we introduce training strategies used in CL-based recommendation methods.
In addition, we summarize the notations used in this survey in Table.~\ref{tab:notations}.
\begin{table*}[t]
  \caption{Key notations.}
  \label{tab:notations}
  \begin{tabular}{c|c}
    \toprule
    Notations&Discriptions\\
    \midrule
    $\mathcal{U}$& The set of users\\
    $\mathcal{I}$& The set of items\\
    $\mathcal{B}$& The mini-batch\\
    $\mathbf{h}, \mathbf{c}, \mathbf{g}, \mathbf{z}$ & The learned representations\\
    $f_\theta$& The encoder to learn representations\\
    $p_\omega$& The pretext decoder\\
    $q_\phi$& The downstream decoder\\
    $\theta$, $\omega$, $\phi$, $\xi$, $\psi$& Learnable parameters\\
    $\lambda, \rho, \epsilon$& The hyperparameter\\
    $\mathcal{T}$& Data-based view generation strategy\\
    $\mathbf{L}$& The location matrix\\
    $\mathbf{A}$ & The adjacent matrix of graph\\
    $\mathbf{X}$ & The feature matrix\\
    $\mathbf{H}$ & The representation matrix\\
    $\mathcal{G}$ & The graph\\
    $\mathcal{V}$/$\mathcal{E}$ & The set of the graph nodes/edges\\
    $s_u$& The interaction sequence of user $u$\\
    $\mathcal{M I}$& Mutual information function\\
    $||$& Concatenation operation\\
    $\circ$& The Hadamard product\\
    $|\cdot|$&The length of a set\\
  \bottomrule
\end{tabular}
\end{table*}

\subsection{Term Definitions}
\subsubsection{Supervised Learning, Unsupervised Learning, and Self-supervised Learning}
Supervised learning refers to a learning paradigm that trains models with manually annotated labels.
In contrast, unsupervised learning indicates the learning paradigm that trains models without using manually annotated labels.
Self-supervised learning can be viewed as a subset of unsupervised learning as it requires no manually annotated labels.
However, unlike other unsupervised learning methods (e.g., clustering) that concentrate on mining data patterns, self-supervised learning aims to generate supervision signals from the data itself, and models are still trained in supervised settings.

\subsubsection{Pretext Tasks Versus Downstream Tasks}
Pretext tasks are pre-designed tasks to be solved by models (e.g., node self-discrimination~\cite{wuSelfsupervisedGraphLearning2021}). By learning the objective functions of the pretext tasks, models learn more generalized representations from unlabeled data, thus benefiting downstream tasks.
Downstream tasks refer to tasks used to evaluate the quality of representations learned by models.
Specifically, in recommender systems, downstream tasks are the recommendation tasks such as sequential recommendation and social recommendation.
In general, solving downstream tasks requires manually annotated labels.
\begin{figure}
\centering\includegraphics[width=0.9\linewidth]{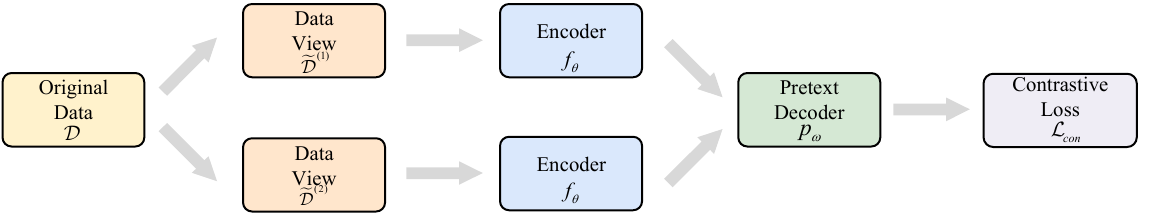}
    \caption{Pipeline of contrastive learning.}
    \label{fig:cl}
\end{figure}

\subsection{Contrastive Learning}
The core idea of contrasting learning (CL) is to maximize agreement between different views, where the agreement is usually measured by Mutual Information (MI). 
The general pipeline of CL is shown in Fig.\ref{fig:cl}. 
In specific, two different data views are generated using view generation strategies. Then, representations in different views are generated by an encoder, which is usually shared by the two views. Finally, the model is optimized by contrastive loss to maximize the agreement between positive pairs and minimize the agreement between negative pairs. 
In general, positive pairs are the same instances from different views, while negative pairs are different instances from different views.
Formally, contrastive self-supervised learning can be formulated as:
\begin{equation} \label{eq:contrastive}
  \theta^{*}, \omega^{*}=\underset{\theta, \omega}{\arg \min } \mathcal{L}_{con}\left(p_\omega\left(f_\theta\left(\tilde{\mathcal{D}}^{(1)}\right), f_\theta\left(\tilde{\mathcal{D}}^{(2)}\right)\right)\right)
\end{equation}
where $\tilde{\mathcal{D}}^{(1)}$ and $\tilde{\mathcal{D}}^{(2)}$ are two generated data views. $f_\theta(\cdot)$ is the (shared) encoder to learn representations of instances in different views. $p_\omega(\cdot)$ is the pretext decoder that estimates the agreement between two instances. 
$\mathcal{L}_{con}$ denotes the contrastive loss.

\begin{figure}[t]
    \centering
    \subfigure[Joint Learning.]{ 
        \includegraphics[width=0.75\linewidth]{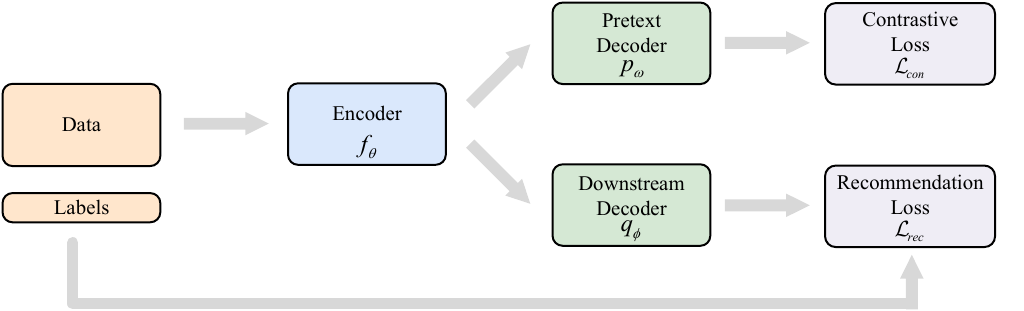}\label{fig:jt}}
    \subfigure[Pre-training and Fine-tuning.]{        
        \includegraphics[width=0.75\linewidth]{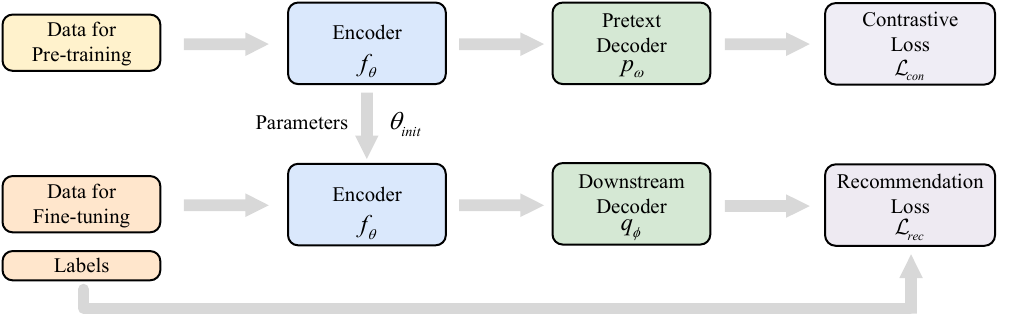}\label{fig:pf}}
    \caption{Two types of training strategies for CL-based recommendation.} 
    \label{fig:training_strategies}
\end{figure}

\subsection{Training Strategy}
Currently, CL-based recommendation methods employ two typical training strategies: Pre-training and Fine-tuning, and Joint Learning. The detailed workflow of them is shown in Fig.~\ref{fig:training_strategies}. 

\subsubsection{Pre-training and Fine-tuning (P\&F)}
In this strategy, the model is trained in two stages. In the pre-training stage, the encoder $f_\theta(\cdot)$ is first pre-trained with contrastive tasks. In addition, the pre-trained parameter $\theta_{init}$ is then used as the initialization parameter for the encoder $f_{\theta_{init}}(\cdot)$. 
In the fine-tuning stage, the pre-trained encoder $f_{\theta_{init}}(\cdot)$ is fine-tuned with the downstream decoder $q_\phi(\cdot)$ supervised by the recommendation task.
The formulation of this strategy can be defined as: 
\begin{equation}
\begin{aligned}
    \theta_{init}, \omega^* &= \arg \min _{\theta, \omega} \mathcal{L}_{con}\left(f_{\theta}, p_\omega\right)\\
    \theta^*, \phi^* &=\arg \min _{\theta_{init}, \phi} \mathcal{L}_{rec}\left(f_{\theta_{init}}, q_\phi\right)
\end{aligned}
\end{equation}
where $\mathcal{L}_{con}$ is the contrastive loss and $\mathcal{L}_{r e c}$ is the recommendation loss. $q_\phi$ is the downstream decoder.

\subsubsection{Joint Learning (JL)}
In this strategy, the encoder $f_\theta(\cdot)$ is jointly trained with the pretext tasks and downstream tasks (i.e., recommendation tasks). Moreover, the encoder is usually shared by pretext and recommendation tasks. This strategy can be considered a type of multi-tasking learning strategy, in which the contrastive pretext task is the auxiliary task to regularize the recommendation task.
The loss function consists of both contrastive loss and recommendation loss. The learning objective can be formalized as:
\begin{equation}
  \theta^{*}, \omega^{*}, \phi^{*} =\arg \min _{\theta, \omega, \phi} \left[\mathcal{L}_{r e c}\left(f_{\theta}, q_\phi\right)+\lambda \mathcal{L}_{con}\left(f_{\theta}, p_{\omega}\right)\right]
\end{equation}
where $\lambda$ is a trade-off hyperparameter that controls the contribution of $\mathcal{L}_{con}$. 

\subsubsection{Discussion}
Compared to JL, P\&F has better generalizability. Specifically, for different recommendation tasks, P\&F only requires fine-tuning, while JL requires re-training. Additionally, when data for the target recommendation task is limited, P\&F allows for pre-training with data from other recommendation tasks. However, due to the two-step training process, P\&F is more complex compared to JL. Moreover, since the pre-trained model is trained without labels, it may not explicitly learn the features of a specific task, which could result in compromised performance on certain tasks.

When recommendation tasks are determined, JL usually achieves better recommendation performance and is easier to be implemented. Therefore, most of the existing CL-based approaches adopt JL. However, JL requires high computational resources since multiple tasks need to be trained simultaneously. Additionally, careful balancing of loss functions for different tasks is necessary to avoid instability or performance degradation. 
To summarize, if the primary goal is to improve recommendation performance on a specific task, it is recommended to choose JL. Conversely, if the goal is to achieve good performance on different recommendation tasks, P\&F is the better choice.

\section{Taxonomy}\label{sec:taxonomy}
In this section, we first propose a unified framework of CL-based recommendation methods. Then we introduce our proposed taxonomy with three perspectives.

\subsection{Unified Framework}\label{sec:unified_framework}
As introduced in Section~\ref{sec:background}, the general framework of CL-based methods is first to perform view generation strategies to obtain multiple views and then maximize the agreement of positive pairs in these views by conducting the contrastive pretext task. Specifically, given the data $\mathcal{D}$, $K$ data views $\{\tilde{\mathcal{D}}^{(k)}\}_{k=1}^{K}$ are obtained through $K$ data-based augmentations $\{\mathcal{T}_k(\cdot)\}_{k=1}^{K}$, which can be formulated as: 
\begin{equation}
    \tilde{\mathcal{D}}^{(k)} = \mathcal{T}_k(\mathcal{D}), k = 1,\cdots,K
\end{equation}
Then, encoders $\{f_{\theta_k}(\cdot)\}_{k=1}^K$ are applied to generate representations $\{\mathbf{h}_k\}_{k=1}^K$ for each data view. Formally, we have 
\begin{equation}
    \mathbf{h}_k = f_{\theta_k} (\tilde{\mathcal{D}}^{(k)}), k = 1,\cdots,K
\end{equation}
In addition, $\{\mathbf{h}_k\}_{k=1}^K$ may have different scales depending on the type of pretext tasks. For example, it can be a representation of an item or a representation of a sequence that consists of multiple items. 

During training, contrastive learning is to maximize the agreement between representations of positive pairs $(\mathbf{h}_i, \mathbf{h}_j)$ in two views. Moreover, the mutual information $\mathcal{M I}\left(\mathbf{h}_i, \mathbf{h}_j\right)$ is usually applied to measure the agreement. The contrastive objective can be defined as:
\begin{equation}\label{eq:objective}
\max _{\left\{\theta\right\}_{i=1}^K} \sum_i \sum_{i \neq j} \lambda_{i j} \mathcal{M I}\left(\mathbf{h}_i, \mathbf{h}_j\right)
\end{equation}
where $\lambda \in \{0, 1\}$, if the mutual information between $\mathbf{h}_i$ and $\mathbf{h}_j$ is calculated then $\lambda=1$, otherwise $\lambda=0$.

Since it is difficult to directly calculate mutual information, mutual information estimators 
are usually used instead. The estimation is calculated based on the discriminator $p_\omega(\cdot)$ (i.e., the pretext decoder). Moreover, projection heads~\cite{CLUE} can be optionally applied to $\{\mathbf{h}_k\}_{k=1}^K$, defined as: 
\begin{equation}
    \mathbf{z}_k = g_{\xi_k}(\mathbf{h}_k), k=1,\cdots,K
\end{equation}
where $g_{\xi_i}(\cdot)$ is a projection head, which can be the Multi-Layer Perceptron (MLP) or linear projection. For the sake of convenience, we treat the projection head as part of the pretext decoder $p_\omega(\cdot)$. 
Then $f_{\theta^*}(\cdot)$ and $p_{\omega^*}(\cdot)$ can be obtained by learning Eq.(\ref{eq:objective}).
Furthermore, by utilizing $f_{\theta^*}(\cdot)$, the generated representations can be used for recommendation tasks. The recommendation task can be formulated as: 
\begin{equation}
    \theta^{* *}, \phi^*=\underset{\theta^*, \phi}{\arg \min } \mathcal{L}_{rec}\left(q_\phi(f_{\theta^*}(\mathcal{D})), y\right)
\end{equation}
where $y$ denotes the labels. $\mathcal{L}_{{rec}}$ is the supervised loss for recommendation tasks such as the cross-entropy (CE) loss.
\begin{figure}[t]
    \centering
    \includegraphics[width=0.9\linewidth]{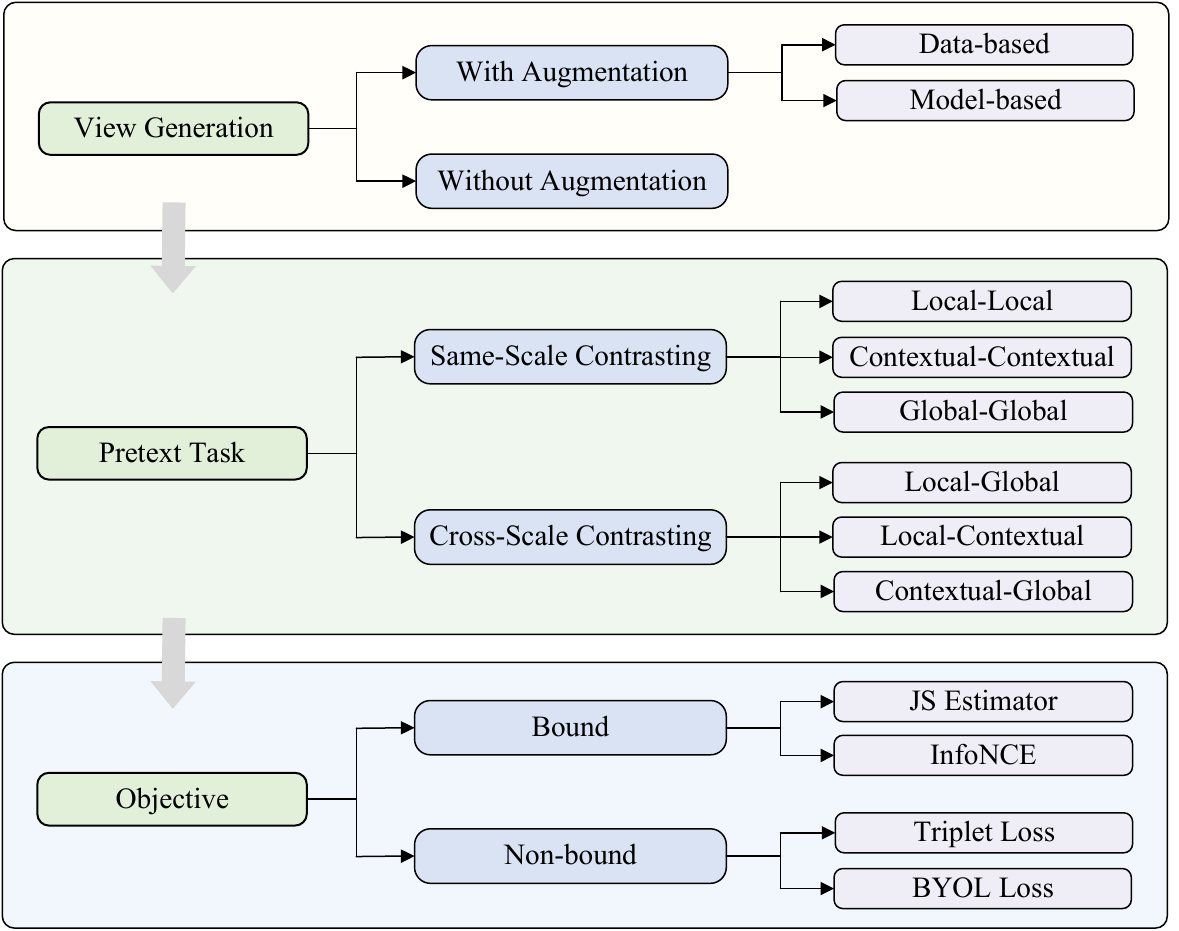}
    \caption{Taxonomy of contrastive learning-based recommendation.
    "JS" refers to Jensen-Shannon. "InfoNCE" loss refers to the loss proposed by~\citet{oord2018representation}, where "NCE" stands for Noise-Contrastive Estimator. "BYOL" (Bootstrap Your Own Latent) loss refers to the contrastive loss proposed by~\citet{BYOL}.}
    \label{fig:taxonomy}
\end{figure}
\subsection{Proposed Taxonomy}
The differences among contrastive learning methods lie in three key components: view generation strategies, pretext tasks, and contrastive objectives. A CL-based recommendation method can be determined by specifying these components. Note that the encoder $f_\theta(\cdot)$ is not included in our taxonomy, as it is not the focus of CL-based recommendation and is determined by the specific recommendation tasks. Therefore, we propose a taxonomy based on these components as shown in Fig.~\ref{fig:taxonomy}. Table.\ref{tab:overview} shows representative works of CL-based recommendation.

{\tiny
\setlength{\LTleft}{0pt} 
\setlength{\LTright}{0pt}
\setlength{\tabcolsep}{4pt}
\begin{longtable} {c|c|c|c|c|c|c|c}
\caption{A summary of CL-based recommendation methods. For alphabets in "Pretext Task", L means Local; C means Contextual; G means Global. For acronyms used, "BPR" refers to Bayesian Personalized Ranking loss; "JL" refers to Joint Learning; "P\&F" refers to Pre-training and Fine-tuning; "CF" refers to Collaborative Filtering; "KG" refers to Knowledge Graph.}\label{tab:overview}\\
    \toprule
    \multicolumn{1}{c|}{Model} & \multicolumn{1}{c|}{Venue} & \multicolumn{1}{c|}{Year}& \multicolumn{1}{c|}{View Generation}& \multicolumn{1}{c|}{Pretext Task}& \multicolumn{1}{c|}{Objective}& \multicolumn{1}{c|}{Training Strategy}& \multicolumn{1}{c}{Recommendation Task}\\ 
    \midrule
    \endfirsthead
    
    \caption[]{A summary of CL-based recommendation methods. (Continued) } \\
    \toprule \multicolumn{1}{c|}{Model} & \multicolumn{1}{c|}{Venue} & \multicolumn{1}{c|}{Year}& \multicolumn{1}{c|}{View Generation}& \multicolumn{1}{c|}{Pretext Task}& \multicolumn{1}{c|}{Objective}& \multicolumn{1}{c|}{Training Strategy}& \multicolumn{1}{c}{Recommendation Task}\\ \midrule
    \endhead
    
    \midrule \multicolumn{8}{r}{{\footnotesize Continued on next page}} \\ 
    \endfoot
    
    \bottomrule
    \endlastfoot

   \multirow{2}{*}{SGL~\cite{wuSelfsupervisedGraphLearning2021}}&  	\multirow{2}{*}{SIGIR}&	\multirow{2}{*}{2021}	&Node/Edge Dropout&		\multirow{2}{*}{L-L}&\multirow{2}{*}{InfoNCE}&	\multirow{2}{*}{JL}&\multirow{2}{*}{Graph-based CF}\\ 
      	& &		&/Random Walk&	&	&	&	\\
    DCL~\cite{DCL}&	Arxiv&	2021&	Edge Dropout&		L-L&InfoNCE&	JL&	Graph-based CF\\
    GDCL~\cite{GDCL}&	DASFAA&	2022&	Graph Diffusion&		L-L&InfoNCE&	JL&	Graph-based CF\\
    SimGCL~\cite{SimGCL}&	SIGIR&	2022&	Embedding Noise&	L-L&InfoNCE	&	JL&	Graph-based CF\\
    XSimGCL~\cite{XSimGCL}&	Arxiv&	2022&	Embedding Noise&	L-L&InfoNCE	&	JL&	Graph-based CF\\
    RocSE~\cite{RocSE}&	TOIS&	2023&	Embedding Noise&		L-L&InfoNCE&	JL&	Graph-based CF\\
    LightGCL~\cite{LightGCL}&	ICLR&	2023&	SVD-based Augmentation	&	L-L&InfoNCE&	JL&	Graph-based CF\\
    RGCF~\cite{RGCF}&	SIGIR&	2022&	Edge Perturbation&	L-L&InfoNCE&	JL&	Graph-based CF\\
    DCCF~\cite{DCCF}&	SIGIR&	2023&	Edge Dropout&	L-L&InfoNCE&	JL&	Graph-based CF\\
    GCARec~\cite{GCARec}&	PKDD&	2022&	Edge Dropout&	L-L&InfoNCE&	JL&	Graph-based CF\\
    LDA\_GCL~\cite{LDA_GCL}&	DASFAA&	2023&	Edge Perturbation&	L-L&InfoNCE&	JL&	Graph-based CF\\
    AdaGCL~\cite{AdaGCL}&	KDD&	2023&	Edge Perturbation&	L-L&InfoNCE&	JL&	Graph-based CF\\
    VGCL~\cite{VGCL}&	SIGIR&	2023&	Model-based Augmentation&	L-L&InfoNCE&	JL&	Graph-based CF\\
    SimRec~\cite{SimRec}&	WWW&	2023&	Model-based Augmentation	&	L-L&InfoNCE&	JL&	Graph-based CF\\
    RGCL~\cite{RGCL}&	SIGIR&	2022&	Node Dropout&		L-L&InfoNCE&	JL&	Review-based\\
    MCLSR~\cite{MCLSR}&	CIKM&	2022&	Without Augmentation&		L-L&InfoNCE	&JL	&Sequential\\
    HCCF~\cite{HCCF}&	SIGIR&	2022&	Edge Dropout&		L-L&InfoNCE	&JL	&Graph-based CF\\
    SGCCL~\cite{li2023sgccl}&	WSDM&	2023&	Edge/Feature Dropout&	L-L&	InfoNCE&	JL&	Graph-based CF\\
    LWC\_KD~\cite{LWC-KD}&	CIKM&	2021&	Model-based Augmentation&	L-L&	InfoNCE&	JL&	Graph-based CF\\
    MPT~\cite{MPT}&	TOIS&	2023&	Node Masking/Substituting/Deleting		&L-L&InfoNCE	&P\&F	&Cold-start\\
    MCCLK~\cite{MCCLK}&	SIGIR&	2022&	Without Augmentation&	L-L&	InfoNCE	&JL	&KG-based\\
    KACL~\cite{KACL}&	WSDM&	2023&	Edge Dropout&	L-L&	InfoNCE	&JL&	KG-based\\
    KGCL~\cite{KGCL}&	SIGIR&	2022&	Edge Dropout&	L-L&	InfoNCE	&JL	&KG-based\\
    KGRec~\cite{KGRec}&	KDD&	2023&	Edge Dropout&	L-L&	InfoNCE	&JL	&KG-based\\
    S-MBRec~\cite{S-MBRec}&	IJCAI&	2022&	Without Augmentation	&	L-L&InfoNCE	&JL&	Multi-behavior\\
    MMCLR~\cite{MMCLR}&	DASFAA&	2022	&Without Augmentation&		L-L	&BPR&JL	&Multi-behavior\\
    KMCLR~\cite{KMCLR}&	WSDM&	2023&	Edge Dropout&	L-L&InfoNCE	&	JL&	Multi-behavior\\
    SEPT~\cite{SEPT}&	KDD&	2022&	Edge Dropout/Predicted amples&		L-L&InfoNCE&	JL&	Social\\
    HGCL\_S~\cite{HGCL_s}&	WSDM&	2023&	Without Augmentation&		L-L	&InfoNCE&JL	&Social\\
    CCDR~\cite{CCDR}&	KDD	&2022&	Subgraph Sampling	&L-L&	InfoNCE&	JL&	Cross-domain\\
    ML-SAT~\cite{ML-SAT}&	CIKM&	2022&	Without Augmentation&	L-L&	InfoNCE	&P\&F&	Cross-domain\\
    DR-MTCDR~\cite{DR_MTCDR}&	TOIS&	2022&	Edge/Node Dropout	&L-L&	InfoNCE&	JL&	Cross-domain\\
    COTREC~\cite{COTREC}&	CIKM&	2022&	Predicted Samples&	L-L&	InfoNCE	&JL&	Session-based\\
    S$^2$-HHGR~\cite{zhangDoubleScaleSelfSupervisedHypergraph2021}&	CIKM&	2021&	Node Dropout&	L-L&	JS&	JL&	Group\\
    SGGCF~\cite{SGGCF}&	WSDM&	2023&	Edge/Node Dropout&	L-L&	InfoNCE	&JL&	Group\\
    CrossCBR~\cite{CrossCBR}&	KDD	&2022&	Edge/Message Dropout &	L-L&	InfoNCE	&JL&	Bundle\\
    DCRec~\cite{DCRec}&	WWW	&2023&	Edge/Message Dropout &	L-L&	InfoNCE	&JL&	Sequential\\
    HMG-CR~\cite{yangHyperMetaPathContrastive2021}&	Arxiv&	2021&	Without Augmentation&	C-C	&	InfoNCE&JL&	Multi-behavior\\
    \multirow{2}{*}{CHEST~\cite{CHEST}}&  	\multirow{2}{*}{TOIS}&	\multirow{2}{*}{2023}	&Subgraph Sampling&		\multirow{2}{*}{C-C}&\multirow{2}{*}{InfoNCE}&	\multirow{2}{*}{P\&F}&\multirow{2}{*}{HIN-based}\\ 
      	& &		&/Path Dropout/Inserting/Substituting&	&	&	&	\\
    KGIC~\cite{KGIC}&	CIKM&	2022&	Subgraph Sampling&	C-C&	InfoNCE	&JL	&KG-based\\
    GCL4SR~\cite{GCL4SR}&	IJCAI&	2022&	Subgraph Sampling&C-C&	InfoNCE	&	JL&	Sequential\\
    MISS~\cite{MISS}&	ICDE&	2022&	Feature Extractor	&C-C&	InfoNCE&	JL&	Sequential\\
    CL4SRec~\cite{CL4SRec}&	ICDE&	2022&	Item Masking/Shuffling/Cropping	&G-G&	InfoNCE&	JL&	Sequential\\
    H$^2$SeqRec~\cite{H2SeqRec}	&CIKM&	2021&	Item Masking/Cropping	&	G-G&InfoNCE&	P\&F&	Sequential\\
    \multirow{2}{*}{CoSeRec~\cite{CoSeRec}}&  	\multirow{2}{*}{Arxiv}&	\multirow{2}{*}{2021}	&Item Shuffling/Cropping &		\multirow{2}{*}{G-G}&\multirow{2}{*}{InfoNCE}&	\multirow{2}{*}{JL}&\multirow{2}{*}{Sequential}\\ 
  & &		&/Masking/Substituting/Inserting&	&	&	&	\\
    ContraRec~\cite{ContraRec}&	TOIS&	2022&	Item Masking/Shuffling/Overlapping&		G-G&InfoNCE&	JL&	Sequential\\
    \multirow{2}{*}{TiCoseRec~\cite{TiCoseRec}}&  	\multirow{2}{*}{AAAI}&	\multirow{2}{*}{2023}	&Ti-crop/Ti-mask/Ti-reorder &		\multirow{2}{*}{G-G}&\multirow{2}{*}{InfoNCE}&	\multirow{2}{*}{JL}&\multirow{2}{*}{Sequential}\\ 
  & &		&/Ti-substitute/Ti-insert&	&	&	&	\\
    \multirow{2}{*}{IOCRec~\cite{IOCRec}}&  	\multirow{2}{*}{WSDM}&	\multirow{2}{*}{2023}	&Item Shuffling/Cropping &	\multirow{2}{*}{G-G}&	\multirow{2}{*}{InfoNCE}&	\multirow{2}{*}{JL}&\multirow{2}{*}{Sequential}\\ 
  & &		&/Masking/Substituting/Inserting&	&	&	&	\\
    \multirow{2}{*}{MCCM~\cite{MCCM}}&  	\multirow{2}{*}{WSDM}&	\multirow{2}{*}{2023}	&Feature Extractor&		\multirow{2}{*}{L-L/G-G}&\multirow{2}{*}{InfoNCE}&	\multirow{2}{*}{JL}&\multirow{2}{*}{News}\\ 
  & &		&/FItem Masking/Instituting&	&	&	&	\\
    CCL~\cite{CCL}&	CIKM&	2021&	Item Masking/Sequence Generator	&G-G&	InfoNCE&	P\&F&	Sequential\\
    MIC~\cite{MIC}	&CIKM&	2022&	Feature/Message Dropout	&	G-G&InfoNCE	&JL	&Sequential\\
    EC4SRec~\cite{EC4SRec}	&CIKM&	2022&	Item Cropping/Masking/Shuffling&G-G&	InfoNCE	&	JL	&Sequential\\
    DuoRec~\cite{DuoRec}&	WSDM&	2022&	Message Dropout&G-G&	InfoNCE	&	JL	&Sequential\\
    CBiT~\cite{CBiT}&	CIKM&	2022&	Item Masking/Message Dropout&G-G	&InfoNCE	&	JL	&Sequential\\
    \multirow{2}{*}{ContrastVAE~\cite{ContrastVAE}}&  	\multirow{2}{*}{CIKM}&	\multirow{2}{*}{2022}	&Item Masking/Cropping/Shuffling&		\multirow{2}{*}{G-G}&\multirow{2}{*}{InfoNCE}&	\multirow{2}{*}{JL}&\multirow{2}{*}{Sequential}\\ 
      & &		&/Message Dropout/Variational Dropout&	&	&	&	\\
    \multirow{2}{*}{CLUE~\cite{CLUE}}&  	\multirow{2}{*}{CIKM}&	\multirow{2}{*}{2022}	&Item Masking/Cropping/Shuffling&		\multirow{2}{*}{G-G}&\multirow{2}{*}{BYOL}&	\multirow{2}{*}{P\&F}&\multirow{2}{*}{Sequential}\\ 
      & &		&/Message Dropout&	&	&	&	\\
      FDSA\_CL~\cite{FDSA_CL} &TKDE&2023&Message Dropout&G-G&InfoNCE&JL&Sequential\\
    \multirow{2}{*}{EMKD~\cite{EMKD}}&  	\multirow{2}{*}{SIGIR}&	\multirow{2}{*}{2023}	&Item Masking&		\multirow{2}{*}{G-G}&\multirow{2}{*}{InfoNCE}&	\multirow{2}{*}{JL}&\multirow{2}{*}{Sequential}\\ 
    & &		&/Model-based Augmentation&	&	&	&	\\
    MCLRec~\cite{MCLRec} &SIGIR&2023&Sequence Augmentor&G-G&InfoNCE&JL&Sequential\\
    DHCN~\cite{DHCN}&	AAAI&	2021&	Feature Shuffling&	G-G	&	JS&JL	&Session-based\\
    OD-Rec~\cite{on_device}	&SIGIR&	2022&	Without Augmentation	&	L-C	&InfoNCE&JL&	Session-based\\
    CGL~\cite{CGL_tois22}&	TOIS&	2022&	Without Augmentation&		G-G	&JS&JL	&Session-based\\
    CFM~\cite{CFM}&	CIKM	&2021&	Feature Dropout	&	G-G&InfoNCE&	JL&	Feature-based\\
    \multirow{2}{*}{CL4CTR~\cite{CL4CTR}}&  	\multirow{2}{*}{WSDM}&	\multirow{2}{*}{2023}	&Message/feature Dropout&	\multirow{2}{*}{G-G}&	\multirow{2}{*}{BYOL}&	\multirow{2}{*}{JL}&\multirow{2}{*}{CTR Prediction}\\ 
  & &		&/Dimension Masking&	&	&	&	\\
    CLCRec~\cite{CLCRec}&	ACM MM&	2021&	Without Augmentation&	G-G&	InfoNCE&	JL&	Cold-start\\
    NCL~\cite{linImprovingGraphCollaborative2022a}&	WWW&	2022	&Clustering&L-C&	InfoNCE		&JL	&Graph-based CF\\
    ICL~\cite{chenIntentContrastiveLearning2022}&	WWW&	2022	&Clustering	&	L-C&InfoNCE&	JL&	Sequential\\
    SITN~\cite{SITN}& AAAI&2023&Clustering&L-C&InfoNCE&P\&F&Cross-domain\\
    MHCN~\cite{MHCN}&	WWW	&2021&	Subgraph Sampling/Feature Shuffling	&	L-C&Triplet&	JL&	Social\\
    SMIN~\cite{longSocialRecommendationSelfSupervised2021a}&	CIKM	&2021&	Graph Diffusion&	L-C&	JS&	JL&	Social\\
CubeRec~\cite{CubeRec}&	SIGIR	&2022&	Without Augmentation&	L-C&	Triplet&	JL&	Group\\
    \multirow{2}{*}{EGLN~\cite{yang2021egln}}&  	\multirow{2}{*}{SIGIR}&	\multirow{2}{*}{2021}	&Edge Dropout/Adding&	\multirow{2}{*}{L-G}&	\multirow{2}{*}{JS}&	\multirow{2}{*}{JL}&\multirow{2}{*}{Graph-based CF}\\ 
      & &		&/Feature Shuffling&	&	&	&	\\
    HGCL~\cite{cai2022hgcl}	&TMM&	2022&	Feature Shuffling&		L-G	&JS&JL	&Micro-video\\
    GroupIM~\cite{GroupIM} &SIGIR&	2020&	Without Augmentation&		L-G	&JS&JL	&Group\\
    BiGI~\cite{cao2021bigi}	&WSDM&	2021&	Subgraph Sampling&C-G&	JS&	JL&Graph-based CF\\
    MMSSL~\cite{MMSSL}&	WWW	&2023&	Without Augmentation&	C-G	&	InfoNCE&JL&	Graph-based CF\\
    C$^{2}$DSR~\cite{C2DSR}&	CIKM&	2022&	Item Substituting&	C-G	&	JS&JL&	Cross-domain\\
    SSI~\cite{SSI}& IJCAI&2021& Item Masking&C-G&InfoNCE&P\&F&Sequential\\
    SESRec~\cite{SESRec}& SIGIR&2023& Without Augmentation&C-G&Triplet&P\&F&Sequential\\
    S$^3$-Rec~\cite{zhouS3RecSelfSupervisedLearning2020}&	SIGIR&	2020&	Item Masking/Cropping&	L-C/C-G/L-G&	InfoNCE	&P\&F&	Sequential\\
    TCPSRec~\cite{TCPSRec}&	CIKM&	2022&	Sequence Dividing	&L-C/L-G/C-C&	InfoNCE&	P\&F&	Sequential\\
\end{longtable}
}

\textbf{View Generation} is the design of how to generate contrastive views. 
Depending on whether the augmentation is needed, we classify the view generation strategies into view generation with augmentation and without augmentation.

\textbf{Pretext Task} is the design of how to obtain supervision signals.
Depending on the scale of the instances being contrasted, we classify the pretext tasks into same-scale contrasting and cross-scale contrasting.

\textbf{Contrastive Objective} is the design of how to measure mutual information.
Depending on whether an estimation of lower-bound of mutual information is provided, we classify the contrastive objectives into bound objective and non-bound objective.

Note that the selection of these components depends on the characteristics of the input data and downstream tasks. For instance, sequence-based augmentation methods may not be suitable for graph data. For sequential recommendation, the pretext tasks generally contrast sequence representations as the primary objective is to learn high-quality sequence representations. It is worth noting that the selection of components is not entirely independent. Although the same pretext task can be performed with different view generation strategies or contrastive objectives, some may not be effective. For instance, feature-based augmentation methods may not be effective when the pretext task aims to model the sequential relationships of items. Therefore, when designing a CL-based recommendation method, we can first design the pretext task based on the specific recommendation task and then select view generation strategies and contrastive objectives accordingly.

\section{View Generation}\label{sec:view_gen}
Recent works~\cite{wang2022strongerAug, tian2020makesview, li2022nlpAug} in other fields have shown that contrastive learning relies heavily on view generation, as generating multiple views facilitates models to explore richer underlying semantic information. In practice, if multiple data views naturally exist, such as interaction views and social networks in social recommendation, pretext tasks can be performed directly on these views. 
In addition, multiple views are not available in many scenarios, so augmentations are needed to generate contrastive views from the original data~\cite{gao2021simcse, chen2020simclr, gutmann2010nce}.
Therefore, we divide existing view generation strategies into view generation with augmentation and without augmentation.

\subsection{With Augmentation}
Augmentation strategies can be categorized into data-based augmentation and model-based augmentation. 
The former generates views based on the data, while the latter is based on the model (i.e., the encoder).

\subsubsection{Data-based Augmentation}
Based on the type of data to be augmented, we classify data-based augmentation into graph-based augmentation, sequence-based augmentation, and feature-based augmentation.
\begin{figure}
    \centering
    \includegraphics[width=0.6\linewidth]{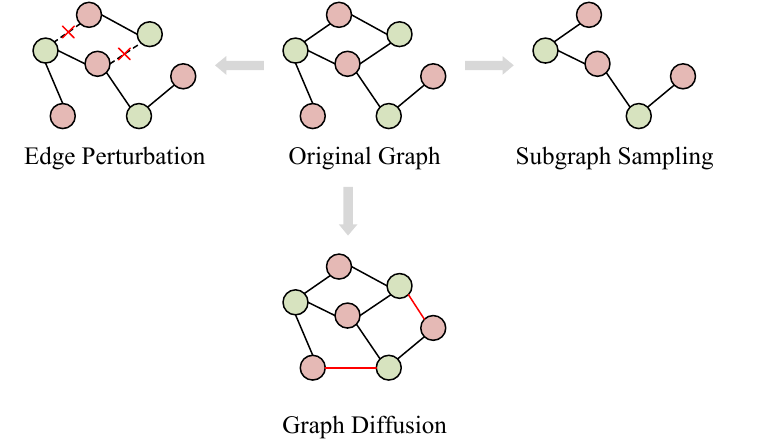}
    \caption{Graph-based augmentation.}
    \label{fig:graph_aug}
\end{figure}

\textbf{Graph-based Augmentation.}
This strategy (shown in Fig.\ref{fig:graph_aug}) performs augmentations on the graph (e.g., interaction graph and social graph) to generate multiple views. Note that since the augmentations of node attributes in graphs are similar to feature-based augmentation, under this subcategory we only present the augmentations of the graph structure (shown in Fig.~\ref{fig:graph_aug}).
Formally, given a graph $\mathcal{G} = (\mathcal{V}, \mathcal{E})$, graph-based augmentation transforms the adjacent matrix $\mathbf{A}$ of $\mathcal{G}$, i.e., $\mathcal{T} = \mathcal{T}(\mathbf{A})$.

\textit{Edge perturbation.}
This strategy~\cite{wuSelfsupervisedGraphLearning2021, DCL, HCCF, li2023sgccl, KGCL, KACL, SGGCF, CrossCBR, DR_MTCDR,yang2021egln, CHEST} generates graph views through randomly adding or dropping edges. It can be defined as:
\begin{equation}
    \mathcal{T}(\mathbf{A}) = \mathbf{A} \circ (1- \mathbf{L}) + (1 - \mathbf{A}) \circ \mathbf{L}
\end{equation}
where $\mathbf{L}$ is the location matrix. If $\mathbf{L}_{ij}=1$, the edge between $i$ and $j$ will be perturbed. Specifically, if $\mathbf{A}_{ij}=1, \mathbf{L}_{ij}=1$, the edge between $i$ and $j$ will be dropped. If $\mathbf{A}_{ij}=0, \mathbf{L}_{ij}=1$, an edge will be added between $i$ and $j$.
$\mathbf{L}$ can be randomly sampled~\cite{wuSelfsupervisedGraphLearning2021, KGCL} or manually set.
Furthermore, $\mathbf{L}$ can also be calculated adaptively~\cite{LDA_GCL, GCARec, AdaGCL} to keep important edges while perturbing possibly unimportant ones.

\textit{Graph Diffusion.} The graph diffusion~\cite{GDCL,longSocialRecommendationSelfSupervised2021a} incorporates the global information to the original graph by creating new edges between nodes. It can be formulated as:
\begin{equation}
\mathcal{T}(\mathbf{A})=\sum_{k=0}^{\infty} \Theta_k \mathbf{T}^k
\end{equation}
where $\Theta_k$ is the weighting coefficient. $\mathbf{T}$ denotes the generalized transition matrix. 
For example, SMIN~\cite{longSocialRecommendationSelfSupervised2021a} generates a substructure-aware adjacent matrix and injects it into the user-item interaction graph.

\textit{Subgraph Sampling.}
This strategy samples a node subset and corresponding edges to generate a subgraph as the data view. Existing methods usually obtain the node subset $\mathcal{V}'$ by uniform sampling, ego-net sampling and knowledge-based sampling.
\textit{Uniform sampling}~\cite{wuSelfsupervisedGraphLearning2021, RGCL, MPT,zhangDoubleScaleSelfSupervisedHypergraph2021, DR_MTCDR} uniformly samples a certain portion of nodes and corresponding edges to augment the views. Node dropout belongs to uniform sampling. For example, SGL~\cite{wuSelfsupervisedGraphLearning2021} randomly drops a portion of nodes, which is denoted as $\mathcal{V}_d$. Therefore, the sampled node subset can be obtained by $\mathcal{V}' = \mathcal{V} - \mathcal{V}_d$.
\textit{Ego-net sampling}~\cite{cao2021bigi} samples the $L$-hop neighbors of each node in a graph, also known as the $L$-ego net.
Therefore, the node subset can be represented as
   $\mathcal{V}' = \left\{j| d(i, j) \leq L \right\}$
, where $d(v_i, v_j)$ is the shortest distance between node $i$ and $j$.
\textit{Knowledge-based sampling}~\cite{MHCN, CHEST} incorporates domain knowledge when sampling subgraph. For example, MHCN~\cite{MHCN} designs three types of triangular motifs based on underlying semantics. Motifs specify high-order relations like "having a mutual friend".

\begin{figure}
    \centering
    \includegraphics[width=0.8\linewidth]{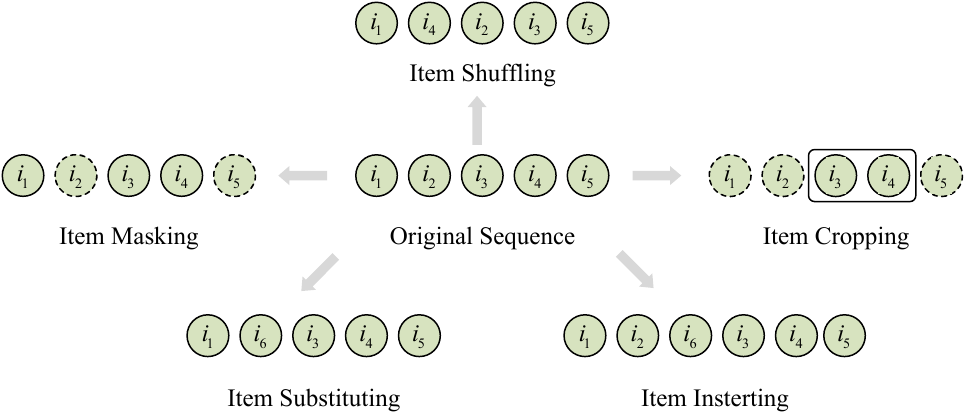}
    \caption{Sequence-based augmentation.}
    \label{fig:seq_aug}
\end{figure}
\textbf{Sequence-based Augmentation.}
This strategy (shown in Fig.\ref{fig:seq_aug}) performs augmentations on the user interaction sequences. Formally, give the interaction sequence $s_u$ of user $u$, it can be formulated as $\tilde{s}_u = \mathcal{T}(s_u)$.

\textit{Item Shuffling.}
The item shuffling~\cite{CL4SRec, CoSeRec,ContraRec, IOCRec,ContrastVAE,EC4SRec} randomly shuffle a continuous subsequence of the interaction sequence to generate the augmented sequence:
\begin{equation}
    \mathcal{T}(s_u) = [i_{u,1}, i_{u,2}, \cdots, \tilde{i}_{u,k}, \cdots, \tilde{i}_{u, k+l_s-1}, \cdots, i_{u,|s_u|}]
\end{equation}
where $[{i}_{u,k}, \cdots, {i}_{u, k+l_s-1}]$ is shuffled as $[\tilde{i}_{u,k}, \cdots, \tilde{i}_{u, k+l_s-1}]$ . $l_c = \lceil \rho_s |s_u| \rceil$  is the length of the subsequence and $\rho_s \in [0,1]$.

\textit{Item Cropping.}
The item cropping~\cite{CL4SRec,H2SeqRec,CoSeRec,IOCRec,ContrastVAE,EC4SRec} randomly chooses a continuous sub-sequence of the interaction sequence and can be represented as:
\begin{equation}
    \mathcal{T}(s_u) = \left[i_{u,k}, i_{u,k+1}, \cdots, i_{u,k+l_c-1}\right] 
\end{equation}
where $l_c = \lceil \rho_c |s_u| \rceil$ is the length of the subsequence and $\rho_c \in [0, 1]$ is the hyperparameter.

\textit{Item Masking.}
This strategy~\cite{CL4SRec,H2SeqRec,CoSeRec,ContraRec,CBiT,IOCRec,ContrastVAE,EC4SRec} randomly chooses a portion of items in the interaction sequence and replaces them with a [mask] token, which can be formulated as:
\begin{equation}
    \mathcal{T}(s_u) = \left[\tilde{i}_{u,1}, \tilde{i}_{u,2}, \cdots, \tilde{i}_{u,|s_u|}\right]
\end{equation}
where $\tilde{i}_{u,k} = [\text{mask}]$ if $i_{u,k}$ is masked, otherwise $\tilde{i}_{u,k}=i_{u,k}$. 

\textit{Item Substituting.}~\cite{CoSeRec,IOCRec} 
As dropout-based augmentation methods such as item masking may exacerbate the problem of data sparsity and cold-start, item substituting and item inserting are proposed. The item substituting randomly replaces a portion of items in the sequence with other items, which can be formulated as:
\begin{equation}
    \mathcal{T} = \left[i_{u,1}, i_{u,2}, \cdots, \tilde{i}_{u,k},\cdots,i_{u,|s_u|}\right]
\end{equation}
where $\tilde{i}_{u,k}$ replaces $i_{u,k}$. Moreover, CoSeRec~\cite{CoSeRec} substitutes items with highly correlated items to maintain the item correlations in the sequences.

\textit{Item Inserting.}~\cite{CoSeRec,IOCRec}
Fewer interactions are recorded in the interaction sequence than the complete behavior of the user, as interaction data from other sources may be missing. Therefore, the comprehensive user preferences and item correlations cannot be captured.
To complete the sequence, CoSeRec~\cite{CoSeRec} proposes the item inserting to generate the augmented sequence.
Firstly, it randomly samples a portion of items in the sequence. Then, items that correlated to sampled items are inserted around them:
\begin{equation}
    \mathcal{T}(s_u) = [i_{u,1}, i_{u,2}, \cdots, \tilde{i}_{u,k}, {i}_{u,k},\cdots, i_{u,|s_u|}]
\end{equation}
where ${i}_{u,k}$ is the sampled item and $\tilde{i}_{u,k}$ is the item related to it.

\begin{figure}
    \centering
    \includegraphics[width=0.6\linewidth]{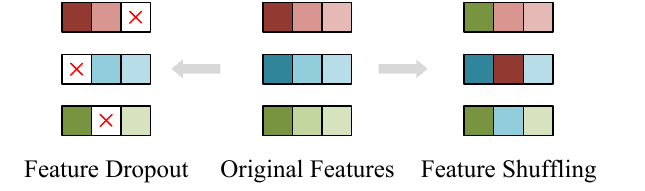}
    \caption{Feature-based Augmentation.}
    \label{fig:feat_aug}
\end{figure}

\textbf{Feature-based Augmentation.}
Feature-based augmentation (shown in Fig.\ref{fig:feat_aug}) performs augmentations on the feature vectors, which can be categorical features or feature representations (e.g., embeddings). Given feature matrix $\mathbf{X}$, the augmented view is represented as $\tilde{\mathbf{X}} = \mathcal{T}(\mathbf{X})$.

\textit{Feature Dropout.}
The feature dropout (masking)~\cite{CFM,MIC,CL4CTR} masks/drops a portion of the features and is formulated as:
\begin{equation}
    \mathcal{T}(\mathbf{X}) = \mathbf{X} \circ (1-\mathbf{L})
\end{equation}
where $\mathbf{L}$ is the masking matrix that indicates the masking locations. If the $j$-th feature of $i$ is masked/dropped, then $\mathbf{L}_{ij}=1$, otherwise $\mathbf{L}_{ij}=0$. Similar to edge perturbation, $\mathbf{L}$ can be uniformly sampled or manually assigned. $\circ$ is the Hadamard product. 

\textit{Feature Shuffling.} The feature shuffling~\cite{yang2021egln, cai2022hgcl, MHCN, DHCN} perturbs the feature matrix by row or column. 
It can be formulated as:
\begin{equation}
    \mathcal{T}(\mathbf{X}) = \mathbf{X}[idx_r, idx_c]
\end{equation}
where $idx_r$ and $idx_c$ are the shuffled row index and the shuffled column index, respectively.

\begin{figure}
    \centering
    \includegraphics[width=\linewidth]{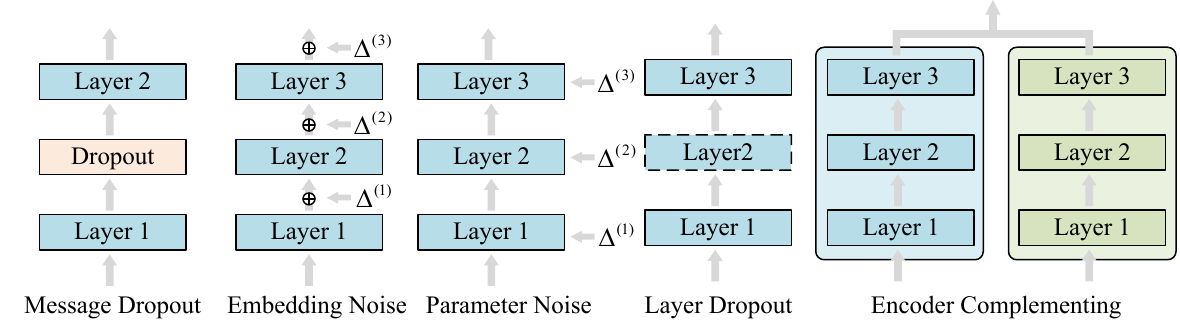}
    \caption{Model-based augmentation.}
    \label{fig:model_aug}
\end{figure}
\subsubsection{Model-based Augmentation}
Model-based augmentation strategies (shown in Fig.\ref{fig:model_aug}) generate views by perturbing the model (i.e., encoder). It is worth noting that unlike data-based augmentation strategies, which first generate different data views and then generate representations for each data view, model-based strategies directly generate different representations for the original data to perform the pretext task. It can be formulated as:
\begin{equation}
    (\mathbf{h}, \mathbf{h}') = (f_{\theta}(\mathcal{D}), f'_{\theta '}(\mathcal{D}))
\end{equation}
where $f_\theta(\cdot)$ and $f'_{\theta '}(\cdot)$ are the encoder and perturbed encoder, respectively. 
$\mathbf{h}, \mathbf{h}'$ are representations output by $f_\theta(\cdot)$ and $f'_{\theta '}(\cdot)$, respectively. 
$\mathcal{D}$ is the original data. 

\textbf{Message Dropout.}
This strategy\cite{DuoRec,CrossCBR,CBiT,CL4CTR,SRMA} randomly masks the neurons in the layers for a certain dropout ratio~\cite{gao2021simcse}. Then, by applying different dropout masks, multiple views can be obtained with the same input data.
For example, DuoRec~\cite{DuoRec} applies two different dropout masks on the Transformer-based model to generate two different views. 

\textbf{Embedding Noise.}
This strategy~\cite{SimGCL,XSimGCL,RocSE} generates different views by adding different noises to original embeddings. 
Unlike feature-based augmentations that only perturb input embeddings or the final representations, this strategy adds noise to the embeddings at different layers of the encoder.
It can be formulated as:
\begin{equation}
    \tilde{\mathbf{E}_l} = \mathbf{E}_l + \Delta_l
\end{equation}
where $\mathbf{E}_l$ is the original embedding and $\Delta_l$ is the perturbation noise at the $l$-th layer. In SimGCL~\cite{SimGCL}, the $\Delta_l \sim U(0,1)$ is the random uniform noise. In RocSE~\cite{RocSE}, $\Delta_l = \epsilon \cdot f_{\text{norm}}(f_{\text{shuffle}}(\mathbf{E}_l))$, where $f_{\text{shuffle}}$ and $f_{\text{norm}}$ are the random shuffling and normalization operations, respectively. $\epsilon$ is a hyper-parameter.

\textbf{Parameter Noise.} This strategy~\cite{xia2022simgrace} adds noises to the parameters of the encoder, which is formulated as:
\begin{equation}
    \theta'_l = \theta_l + \epsilon \Delta_l
\end{equation}
where $\theta_l$ and $\theta'_l$ are the original parameters and perturbed parameters of $l$-th layer, respectively. The $\Delta_l$ is the random noise, that can be sampled from the Gaussian distribution. $\epsilon$ is a hyper-parameter.

\textbf{Architecture Perturbation.}
Unlike the above strategies that perturb learnable parameters in the model, some works generate different views by changing the model architecture. For example, SRMA~\cite{SRMA} proposes \textit{Layer Dropout} and \textit{Encoder Complementing}. Specifically, the \textit{Layer Dropout} randomly drops a portion of layers in the model during training to enable contrastive learning between shallow features and deep features. The \textit{Encoder Complementing} uses a pre-trained encoder to generate representations. These representations are combined with the representations generated by the original encoder for contrastive learning. 
MA-GCL~\cite{MA-GCL} proposes to perturb the architecture of graph neural network (GNN) encoders by varying the number and permutations of propagation and transformation operators.

\subsection{Without Augmentation}
The key idea of contrastive learning is to maximize the agreement between different views. Thus, if multiple views naturally exist, these views can be contrasted directly without additional augmentations. 
For example, in cross-domain recommendation, the two domains can be considered as two views. Therefore, some methods such as CCDR~\cite{CCDR} and ML-SAT~\cite{ML-SAT}, directly perform contrastive learning between these domains.
For knowledge graph-based recommendation, Some methods~\cite{MCCLK,KACL} use the knowledge graph as a contrastive view.
For multi-behavior recommendation, views can be constructed based on the auxiliary behavior data. For example, S-MBRec~\cite{S-MBRec} treats each type of behavior as a view. 
Specifically, HMG-CR~\cite{yangHyperMetaPathContrastive2021} build different hyper meta-graphs based on the hyper meta-paths constructed using the distance between auxiliary behavior and target behavior \textit{buy}. 
In bundle recommendation, user-item interaction and user-bundle interaction can also be contrasted~\cite{CrossCBR}. 

\begin{table*}
\caption{Comparison between different view generation strategies.}
\label{tab:aug_discussion}
\begin{tabular}{cccc}
\toprule
\multicolumn{1}{c|}{\multirow{2}{*}{}} & \multicolumn{2}{c|}{With Augmentation}                                 & \multirow{2}{*}{Without Augmentation} \\ \cline{2-3}
\multicolumn{1}{c|}{}                  & \multicolumn{1}{c|}{Data-based} & \multicolumn{1}{c|}{Model-based} & \\ 
\midrule
Trial-and-errors Free& \XSolidBrush& \CheckmarkBold& \CheckmarkBold\\ 
Domain Knowledge Free& \CheckmarkBold& \CheckmarkBold& \XSolidBrush\\ 
Generalizability& \XSolidBrush& \CheckmarkBold& \XSolidBrush\\
Semantic Preservation& \XSolidBrush& \CheckmarkBold& \CheckmarkBold\\
\bottomrule
\end{tabular}
\end{table*}

\subsection{Discussion}
Table.~\ref{tab:aug_discussion} shows the comparison between different view generation strategies. 
In specific, most existing CL-based recommendation methods adopt data-based augmentation strategies due to their ease of implementation. However, data-based augmentations are usually selected by manual trial-and-errors, which significantly limits the generalizability of these methods. 
In addition, some data-based augmentations destroy the semantic information of the original data, potentially harming recommendation performance~\cite{SimGCL}.

Strategies without augmentation do not require trial-and-errors. These strategies typically use domain knowledge to build auxiliary views, which preserves the semantics of the data. However, domain knowledge is expensive and cannot be applied to other domains. Furthermore, since the views are fixed during model training, strategies without augmentation lack the introduction of randomness that helps to learn noise-invariant representations.

Compared to other strategies, model-based augmentations have better generalizability because they vary the learned representations without considering the original data. Although model-based augmentations require no trial-and-error and domain knowledge, settings such as the dropout ratio of messages/layers still require manual tuning. This limits their generalizability to some extent. Additionally, designing architecture-based perturbations is challenging.

Furthermore, many works~\cite{CBiT, ContrastVAE, CrossCBR, CL4CTR} adopt hybrid methods by combining multiple view generation strategies.
In this way, the advantages of different strategies can be combined. However, some disadvantages may still exist. For instance, combining strategies without augmentation with data-based augmentations can be helpful in introducing randomness but data-based augmentation still requires manual trial-and-errors.
 
To summarize, selecting the appropriate strategy for view generation requires considering various factors. Here, we provide guidance for strategy selection based on typical issues in recommender systems. For the cold-start problem, strategies that drop data should be avoided as it exacerbates the problem~\cite{CoSeRec}. Instead, views can be generated by adding/substituting interactions or using model-based strategies. Incorporating other data, like knowledge graphs or data from different recommendation domains, can also help mitigate data sparsity. To tackle the noise issue, data-based augmentation strategies are often more effective than other strategies as the noise mainly exists in the data. Perturbing data can make models more robust, thus mitigating the impact of noise~\cite{denoise_survey}. Additionally, constructing a denoised data view based on metrics such as edge reliability degrees~\cite{RGCF} can be helpful. Similarly, for addressing bias, a debiased view can also be constructed. Introducing other side information is also beneficial in reducing bias~\cite{Bias_Survey}.

\section{Pretext Task}~\label{sec:pretext_task}
The goal of contrastive learning is to maximize the agreement between positive pairs (i.e., instances with same semantic information) and minimize the agreement between negative pairs (i.e., instances with unrelated semantic information). According to the scale of instances, we classify existing contrastive pretext tasks into two categories: same-scale contrasting and cross-scale contrasting. 

Specifically, there are three contrastive scales: local, contextual, and global. The local scale usually represents the minimum granularity of the input data, while the global scale represents the maximum. For instance, in graph (sequence) data, the local scale represents the node (item/feature), and the global scale represents the whole graph (sequence). The contextual scale is between the local and global scales and represents the subgraph (subsequence).

\begin{figure}
    \centering
    \includegraphics[width=0.4\linewidth]{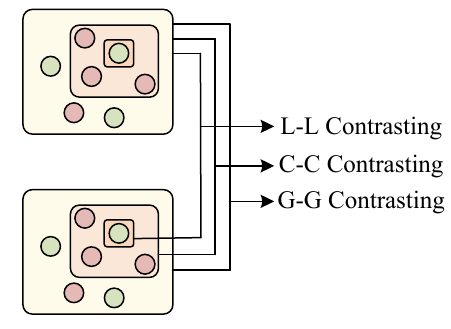}
    \caption{Illustration of same-scale contrasting.}
    \label{fig:same_scale_whole}
\end{figure}
\subsection{Same-Scale Contrasting}
Depending on the different scales being contrasted, same-scale contrasting (shown in Fig.\ref{fig:same_scale_whole}) can be further divided into three sub-types: local-local (L-L) contrasting, contextual-contextual (C-C) contrasting, and global-global (G-G) contrasting. Considering the unique characteristics of the recommendation tasks, we present existing methods based on their recommendation tasks.

\subsubsection{Local-Local Contrasting}
Methods under this category mainly discriminate the local representations (i.e., representation of users/items) and can be formulated as
\begin{equation}
   \theta^{*}, \omega^{*}=\underset{\theta, \omega}{\arg \min } \mathcal{L}_{con}\left(p_\omega\left(\mathbf{h}_i, \mathbf{h}_j\right)\right)
\end{equation}
where $\mathbf{h}_i$ and $\mathbf{h}_j$ are the representation of instance $i$ and $j$ in different views respectively. Furthermore, these representations are generated by encoder $f_\theta(\cdot)$, which is usually shared by different views.

\textbf{Graph-based Collaborative Filtering.}
Depending on the types of graphs being contrasted, methods can be categorized into \textit{contrasting on user-item graph} and \textit{contrasting on different graphs}.

(i) \textit{Contrasting on User-Item Graph.}
As only one graph exists, methods under this category should perform augmentations on the user-item interaction graph to generate different views.

\textbf{SGL}~\cite{wuSelfsupervisedGraphLearning2021} first applies contrastive learning to graph-based recommendation. Given a user-item interaction graph $\mathcal{G}$. It first generates two different graph views $\tilde{\mathcal{G}}^{(1)}=\mathcal{T}(\mathcal{G})$ and $\tilde{\mathcal{G}}^{(2)}=\mathcal{T}(\mathcal{G})$. $\mathcal{T}$ is the data-based view generation strategy. Moreover, it utilizes three data-based augmentations including node dropout, edge dropout, and random walk (apply edge dropout at each layer). 

Then, it utilizes LightGCN~\cite{he2020lightgcn} as graph encoder $f_\theta(\cdot)$ to generate node representations  $\mathbf{H}^{(1)}=f_\theta(\tilde{\mathcal{G}}^{(1)})$ and $\mathbf{H}^{(2)}=f_\theta(\tilde{\mathcal{G}}^{(2)})$. 
Afterward, it performs the node self-discrimination task. Specifically, it makes the representations of the same node (i.e., the positive pair) in different views similar while making representations of different nodes (i.e., the negative pairs) in different views dissimilar. The contrastive loss of the user side can be formulated as: 
\begin{equation}\label{eq:sgl_loss}
    \mathcal{L}^{user}_{con} = - \log \frac{\exp(p_\omega(\mathbf{h}_u^{(1)}, \mathbf{h}_u^{(2)}))}{\exp(p_\omega(\mathbf{h}_u^{(1)}, \mathbf{h}_u^{(2)})) + Neg}
\end{equation}
where $\mathbf{h}_u^{(1)} \in \mathbf{H}^{(1)}$ and $\mathbf{h}_u^{(2)} \in \mathbf{H}^{(2)}$ are representations of user $u$. $p_\omega(\cdot)$ is the cosine similarity with a temperature parameter $\tau$. $p_\omega(\mathbf{z}_u^{(1)}, \mathbf{z}_u^{(2)}) = (\mathbf{z}_u^{(1)})^T\mathbf{z}_u^{(2)}/\tau$ and $\mathbf{z}_u^{(1)} = \mathbf{h}_u^{(1)}/||\mathbf{h}_u^{(1)}||$. 
In addition, for efficiency, the in-batch negative sampling can be adopted, i.e., considering only different nodes of the same batch $\mathcal{B}$ instead of using all other nodes as negative samples. Therefore, $Neg$ is defined as
\begin{equation}\label{eq:neg}
    Neg = \sum_{v\in\mathcal{B}, v \neq u} \exp(p_\omega(\mathbf{h}_u^{(1)}, \mathbf{h}_v^{(2)}))
\end{equation}
Note that $(\mathbf{h}_u^{(1)}, \mathbf{h}_v^{(2)})$ is the \textit{inter-view} negative pairs. The loss of the item side $\mathcal{L}^{item}_{con}$ can be obtained in the same way. The contrastive loss is $ \mathcal{L}_{con} = \mathcal{L}^{user}_{con} + \mathcal{L}^{item}_{con}$.
Finally, SGL adopts a joint learning strategy to optimize the contrastive loss and recommendation loss.

Based on the framework of SGL, several works are proposed.
The main difference with SGL is in the view generation strategies.
\textbf{DCL}~\cite{DCL} perturbs the edges in $L$-ego net of each node to obtain views. 
\textbf{GDCL}~\cite{GDCL} generate new graph view using graph diffusion. Moreover, it constructs the \textit{intra-view} negative pairs and the Eq.(\ref{eq:neg}) can be rewritten as
\begin{equation}
    Neg = \sum_{v\in\mathcal{U}, v \neq u} \exp(p_\omega(\mathbf{h}_u^{(1)}, \mathbf{h}_v^{(2)}) + p_\omega(\mathbf{h}_u^{(1)}, \mathbf{h}_v^{(1)}))
\end{equation}
\textbf{LightGCL}~\cite{LightGCL} proposes a singular value decomposition (SVD)-based graph augmentation strategy to effectively distill global collaborative signals. 
In specific, SVD is first performed on the adjacency matrix. Then, the list of singular values is truncated to retain the largest $K$ values and truncated matrices are used to reconstruct the adjacency matrix. The node contrastive learning is performed between the reconstructed graph and the original graph. 
\textbf{RGCL}~\cite{RGCL} also performs edge contrastive learning. It maximizes the MI between the review representation and the corresponding interaction representation. 

\textbf{SimGCL}~\cite{SimGCL}, \textbf{XSimGCL}~\cite{XSimGCL}, and \textbf{RocSE}\cite{RocSE}
generate views by adding uniform noises to node representations. Moreover, to reduce the computational complexity, XSimGCL~\cite{XSimGCL} replaces the final-layer contrast with cross-layer contrasting. It only utilizes one Convolutional Network (GCN)-based encoder and contrasts embeddings of different layers:
\begin{equation}
\begin{aligned}
\mathcal{L}_{con}= -\log \frac{\exp \left(\mathbf{h}_i^{\top} \mathbf{h}_i^{l^*} / \tau\right)}{\sum_{j \in \mathcal{B}} \exp \left(\mathbf{h}_i^{\top} \mathbf{h}_j^{l^*} / \tau\right)}
\end{aligned}
\end{equation}
where $\mathbf{h}$ is the node representation and $\mathbf{h}^{l^*}$ is the representation at the $l^*$ layer.

\textbf{SimRec}~\cite{SimRec} proposes contrastive knowledge distillation by incorporating contrastive learning into knowledge distillation. It adopts a  GCN-based encoder as the teacher model and an MLP-based model as the student model to generate node (user/item) representations. Furthermore, it maximizes the MI between the representations of the same node learned from the teacher model and the student model.

\textbf{RGCF}~\cite{RGCF} constructs contrastive views based on the edge reliability degree. It first obtains the node structural feature by aggregating its one-hop neighbor representations. Then reliability degree is calculated based on the similarity of the structural feature
\begin{equation}
\begin{gathered}
    \cos \left(\mathbf{h}_u^s, \mathbf{h}_i^s\right)=\frac{\mathbf{h}_u^{s \top} \mathbf{h}_i^s}{\left\|\mathbf{h}_u^s\right\|_2 \cdot\left\|\mathbf{h}_i^s\right\|_2}\\
    s_{u, i}=\left(\cos \left(\mathbf{h}_u^s, \mathbf{h}_i^s\right)+1\right) / 2
\end{gathered}
\end{equation}
where $\mathbf{h}_u^s$ and $\mathbf{h}_i^s$ are the structure feature of user $u$ and item $i$ in the user-item interaction graph. $s_{u,i}$ is the reliability degree of the edge between $u$ and $i$. 
Then, RGCF constructs a denoised graph and a diversity graph. 
The denoised graph is constructed by dropping edges with lower reliability degrees while the diversity graph is constructed by randomly adding edges with higher degrees. It maximizes the MI between representations of the same user in the two graphs.

\textbf{DCCF}~\cite{DCCF} constructs two relation graphs to perform contrastive tasks.
Specifically, it first obtains general representation $\mathbf{h}^z$ and intent-aware representation $\mathbf{h}^r$ for each node. Then two graph relation matrices $\mathcal{G}^z$ and $\mathcal{G}^r$ are generated using them. The calculation of $\mathcal{G}^z$ can be formulated as:
\begin{equation}
\begin{gathered}
\mathcal{M}^z_{u, i}=\left(\cos\left(\mathbf{h}^z_u, \mathbf{h}^z_i\right)+1\right) / 2\\
\mathcal{G}^z=\mathcal{M}^z \circ \mathbf{A}
\end{gathered}
\end{equation}
where $\mathbf{A}$ is the original user-item interaction graph. 
$\mathcal{G}^r$ can be obtained similarly.
The node self-discrimination task is performed on these three graphs. Furthermore, the augmented graph views become learnable in this process. \textbf{GCARec}~\cite{GCARec}, \textbf{LDA$\_$GCL}~\cite{LDA_GCL}, and \textbf{AdaGCL}~\cite{AdaGCL} also use a learnable strategy for generating graph views. Specifically, GCARec applies an MLP to obtain preserving probabilities for edges and uses them to generate graph views on these probabilities by sampling edges. LDA\_GCL also generates views by obtaining probabilities but requires pre-trained models to generate representations for calculating probabilities. AdaGCL utilizes a generative model and a denoised model to generate graph views. 

Similar to ADaGCL, \textbf{VGCL}~\cite{VGCL} generates graph views based on a generative model, but instead of generating graphs, it directly obtains node representations for contrastive learning. VGCL also maximizes the MI between the representations of nodes in the same cluster.

(ii) \textit{Contrasting on Different Graphs.}
In addition to the user-item interaction graph, some works construct other graphs using interaction data. 
That is, views are usually generated without augmentation.

\textbf{MCLSR}~\cite{MCLSR} constructs three graphs based on the interaction sequences, including a user-item relation graph, an item-item relation graph, and a user-user relation graph. 
\textbf{HCCF}~\cite{HCCF} constructs two views, including a user-item interaction graph and a learnable hypergraph. The node self-discrimination same as SGL is performed in MCLSR and HCCF.
\textbf{SGCCL}~\cite{li2023sgccl} constructs a user-user graph $\mathcal{G}_{uu}$ and an item-item graph $\mathcal{G}_{ii}$ and performs edge/feature dropout to augment them. Node self-discrimination is conducted on the $\tilde{\mathcal{G}}_{uu}$ and $\tilde{\mathcal{G}}_{ii}$.  

\textbf{LWC\_KD}~\cite{LWC-KD} incorporates contrastive knowledge distillation into incremental learning. In each time block, a user-item graph, user-user graph, and item-item graph are constructed. It proposes layer-wise structure-aware contrastive learning, which contrasts node representations of the same layer $k$ between adjacent time blocks. It can be formulated as: 
\begin{equation}
\mathcal{L}_{con} = \frac{-1}{\left|\mathcal{N}_{i}^{t-1}\right|} \sum_{i^{\prime} \in \mathcal{N}_{i}^{t-1}} \log \frac{\exp \left(\mathbf{h}_{i, k}^{t-1} \cdot \mathbf{h}_{i^{\prime}, k}^t / \tau\right)}{\exp \left(\mathbf{h}_{i, k}^{t-1} \cdot \mathbf{h}_{i^{\prime}, k}^t / \tau\right) + \sum_{\hat{i} \in {Neg}^{t-1}} \exp \left(\mathbf{h}_{i, k}^{t-1} \cdot \mathbf{h}_{\hat{i}, k}^t / \tau\right)}
\end{equation}
where $t$ denotes the time block. $\mathbf{h}_{i,k}^{t-1}$ is the representation of node $i$. $\mathcal{N}_{i}^{t-1}$ denotes the one-hop neighbors of $i$ at $t-1$. ${Neg}^{t-1}$ denotes nodes randomly selected from the unconnected nodes.
\textbf{MPT}~\cite{MPT} extends PT-GNN~\cite{PT_GNN} which performs reconstruction tasks and can only model the intra-correlations. It leverages contrastive tasks to capture the inter-correlations within the data.
Specifically, it samples subgraphs/paths for each user. 
Node dropout/substitution is applied to augment subgraphs/paths.  
The MI between representations of the same user in augmented subgraphs/paths is maximized. 

\textbf{Knowledge Graph-based Recommendation.}
Apart from interaction data, the knowledge graph (KG) is also utilized for CL-based recommendation, as it can bring rich semantic information.

Generally, CL-based recommendation methods using KG generate views by manual design.
\textbf{MCCLK}~\cite{MCCLK} constructs three graph views, including user-item graph $\mathcal{G}_{ui}$, item-entity graph $\mathcal{G}_{ie}$ and user-item-entity graph $\mathcal{G}_{uie}$. 
It maximizes the MI between the user representations $\mathbf{h}_u^{(ui)}$, $\mathbf{h}_u^{(uie)}$ in $\mathcal{G}_{ui}$ and $\mathcal{G}_{uie}$ and between the item representations $\mathbf{h}_i^{(ui)}$, $\mathbf{h}_i^{(ie)}$ in $\mathcal{G}_{ui}$ and $\mathcal{G}_{ie}$. Furthermore, it generates a new representation for each item
\begin{equation}
\mathbf{h}_i^\prime = \mathbf{h}_i^{(ui)} || \mathbf{h}_i^{(ie)}    
\end{equation}
where $||$ is the concatenation operation. 
Then the MI between $\mathbf{h}_i^\prime$ and $\mathbf{h}_i^{(uie)}$ is maximized by performing node-self discrimination.
\textbf{KACL}~\cite{KACL} maximizes the MI between representations of the same item in the augmented user-item graph and the augmented knowledge graph. Moreover, the augmented graphs are generated by automatically dropping unimportant edges. 

KG can also be used to guide the generation of different user-item graph views. 
For example, \textbf{KGCL}~\cite{KGCL} performs stochastic augmentation on the knowledge graph to generate two different views $\tilde{\mathcal{G}}_k^{(1)}$ and $\tilde{\mathcal{G}}_k^{(2)}$. 
The the item consistency is $c_i= \cos (\mathbf{h}^{(1)}_i, \mathbf{h}^{(2)}_i)$. 
$\cos (\cdot)$ is the cosine similarity function. $\mathbf{h}^{(1)}_i$ and $\mathbf{h}^{(2)}_i$ are item representations in $\tilde{\mathcal{G}}_k^{(1)}$ and $\tilde{\mathcal{G}}_k^{(2)}$, respectively.
Then, the user-item interaction graph $\mathcal{G}_{ui}$ is augmented using knowledge-guided data augmentation. Specifically, edge dropout is performed based on the probability calculated as follows: 
\begin{equation}
\begin{gathered}
w_{ui}  =\exp \left(c_i\right)\\
p_{ui}^{\prime} =\max \left(\frac{w_{ui}-w^{\max }}{w^{\max }-w^{\min }}, p_\tau\right) \\
p_{ui}  =p_a \cdot \mu_{p^{\prime}} \cdot p_{ui}^{\prime}
\end{gathered}
\end{equation}
where $p_{ui}$ is the dropout probability of edge $(u,i)$ in $\mathcal{G}_{ui}$. $p_\tau$ is the threshold. 
$p_{ui}^{\prime}$ is an intermediate variable that is integrated with the mean value $\mu_{p^{\prime}}$.
$p_a$ is a strength controller. With the $p_{ui}$, masking vectors $\mathcal{M}_i \in \{0,1\}$ are generated based on the Bernoulli distribution~\cite{marshall1985family}. The augmented graphs is $\tilde{\mathcal{G}}_{ui}^{(i)} = (\mathcal{V}, \mathcal{E} \circ \mathcal{M}_i)$. 
Moreover, KGCL performs both intra-view contrasting and inter-view contrasting. \textbf{KGRec}~\cite{KGRec} generates rationale scores of knowledge triplets based on attention mechanism to augment the KG and user-item graph. Besides, it only performs inter-view contrasting.

\textbf{Multi-behavior Recommendation.}
For enhancing user intention modeling, multi-behavior recommendation methods incorporate multiple types of user behavior data, which can be used to build contrastive views.
\textbf{S-MBRec}~\cite{S-MBRec} adopts a star-style contrastive task, i.e., it only performs contrastive learning between the target behavior (usually the buy) and each auxiliary behavior. It samples positive samples based on the similarity under target behavior. The similarity is calculated by point-wise mutual information~\cite{PMI}. 
\begin{equation}
\begin{gathered}
P M I\left(u, u^{\prime}\right)  =\log \frac{p\left(u, u^{\prime}\right)}{p(u) p\left(u^{\prime}\right)}, \\
p(u) =\frac{|\mathcal{I}(u)|}{|\mathcal{I}|}, \\
p\left(u, u^{\prime}\right) =\frac{\left|\mathcal{I}(u) \cap \mathcal{I}\left(u^{\prime}\right)\right|}{|\mathcal{I}|},
\end{gathered}
\end{equation}
where $\mathcal{I}(u)$ is items that user $u$ has interacted. $\mathcal{I}$ is the item set and $|\mathcal{I}|$ is the number of items.
If similarity $P M I\left(u, u^{\prime}\right) > t$, $(u, u^{\prime})$ are considered as positive pairs. $t$ is the threshold. The positive samples of the items are selected in a similar way. Moreover, negative samples are selected randomly. 

\textbf{MMCLR}~\cite{MMCLR} constructs a graph view (user-item graph) $\mathcal{G}$ and a sequence view (multi-behavior sequence) $\mathcal{S}$. For each view, different behavior representations of the same users (e.g., $\mathbf{h}_{u,b_1}^{g}$ and $\mathbf{h}_{u,b_2}^{g}$) are treated as positive pairs. It also maximizes the MI between overall representations of the same user (e.g., $\mathbf{h}_{u}^g$ and $\mathbf{h}_{u}^s$) in different views. 
\textbf{KMCLR}~\cite{KMCLR} maximizes the MI between different behaviors of the same user. In addition, it performs knowledge-aware contrastive learning. It leverages a knowledge graph to guide the augmentation of the user-item graph under the target behavior $\mathcal{G}_{ui}^{b_t}$. 
It first calculates the consistency $c_i$ of each item $i$ like KGCL.
Then the edge dropout probability is obtained by
\begin{equation}
\begin{gathered}
\hat{p}_{ui}=\sigma\left(\mathbf{h}_u^T \mathbf{h}_i\right) \circ c_i, \\
p_{ui}=\left(1-\operatorname{Min} \_\operatorname{Max}\left(\hat{p}_{u i}\right)\right) a+\operatorname{Min\_ Max}\left(\hat{p}_{ui}\right) b
\end{gathered}
\end{equation}
where $\mathbf{h}_u$ and $\mathbf{h}_i$ are the user representation and item representation in $\mathcal{G}_{ui}^{b_t}$, respectively. $\operatorname{Min} \_\operatorname{Max}(\cdot)$ is the min-max normalization
function. $a$ and $b$ are hyperparameters that control the value interval of $p_{ui}$.
Then it performs edge dropout in a way similar to KGCL. Moreover, KMCLR adopts the node self-discrimination task.

\textbf{Social Recommendation.}
For social recommendation, social networks are utilized to improve recommendation performance. 
Similar to KG-based recommendation, views can be generated based on manual design.

\textbf{SEPT}~\cite{SEPT} constructs three graph views based on the interaction data and social network, including a preference view $\mathcal{G}_r$, friend view $\mathcal{G}_f$, and sharing view $\mathcal{G}_s$. 
$\mathcal{G}_r$ is the user-item interaction graph. Other views are constructed based on two types of triangle motifs. 
Moreover, it leverages tri-training~\cite{zhou2005tri} to predict positive samples for each view. The Eq.(\ref{eq:sgl_loss}) is changed as follows
\begin{equation}
\mathcal{L}_{con}=- \sum_{v \in\{r, s, f\}}\log \frac{\sum_{p \in \mathcal{P}_{u+}^v} p_\omega\left(\mathbf{h}_u^v, \tilde{\mathbf{h}}_p\right)}{\sum_{p \in \mathcal{P}_{u+}^v} p_\omega\left(\mathbf{h}_u^v, \tilde{\mathbf{h}}_p\right)+\sum_{j \in \mathcal{U} / \mathcal{P}_{u+}^v} p_\omega\left(\mathbf{h}_u^v, \tilde{\mathbf{h}}_j\right)}
\end{equation}
where $\mathcal{P}_{u+}^v$ is the set of predict positive samples.
$\mathbf{h}_u^{v}$ is the user representation in view $v$. 
$p_\omega$ is the cosine similarity with temperature parameter $\tau$. $\tilde{\mathbf{h}}_p$ is the representation of user $p$ in $\tilde{\mathcal{G}}$, which is obtained by performing random edge dropout on the joint graph of the user-item interaction graph and the social network.

\textbf{HGCL\_S}~\cite{HGCL_s} constructs three types of graphs: user-item graph, user-user graph, and item-item graph. It performs the node self-discrimination task on the corresponding graphs. Moreover, when generating node representations in the user-user graph and item-item graph, HGCL\_S utilizes a meta-network.

\textbf{Cross-domain Recommendation.}
Different domains in cross-domain recommendation can be considered as different views.
For cross-domain recommendation, two types of contrastive tasks can be conducted, i.e., \textit{single-domain} contrasting and \textit{cross-domain} contrasting. 

\textbf{CCDR}~\cite{CCDR} performs both  single-domain contrasting and cross-domain contrasting. For the single-domain contrasting, it samples two subgraphs of each node to generate two node representations. Then the MI between these two representations is maximized. 
For the cross-domain contrasting,
It maximizes the MI between the representations of the same node in the source domain and the target domain. Besides, to extract more cross-domain knowledge between unaligned nodes, CCDR maximizes the MI between the representation of an aligned node in the source domain and the representations of its target-domain neighbors.

\textbf{ML-SAT}~\cite{ML-SAT} studies the multi-scenario problem, which can be viewed as a multi-domain recommendation problem. Moreover, it only performs the cross-domain contrasting between two different scenarios. It treats the representations of the same users/items in different domains as positive pairs and representations of other users/items in both scenarios as negative pairs.

\textbf{DR-MTCDR}~\cite{DR_MTCDR} only performs sing-domain contrasting. It augments the user-item graph in each domain by edge/node dropout. For each domain, it generates $K$ channel node representations. For each channel, MI between representations of the same node is maximized. 

\textbf{Group Recommendation.}
Group recommendation aims to recommend items that can fulfill the preferences of a collective of users. 
To improve group recommendation performance, \textbf{S$^2$-HHGR}~\cite{zhangDoubleScaleSelfSupervisedHypergraph2021} builds a hierarchical hypergraph based on the user-item, group-item, and user-group interactions. It applies double-scale node dropout strategies including coarse- and fine-grained dropout on the hypergraph. The former drops users in all groups. The latter only drops some nodes in a specific group while other groups still contain these users. It performs node self-discrimination on the coarse- and fine-grained user representations as follows: 
\begin{equation}
\begin{gathered}
\mathcal{L}_{con}=-\log \sigma\left(p_\omega\left(\mathbf{h}_u^{\prime}, \mathbf{h}_u^{\prime \prime}\right)\right)-\sum_{j=1}^n\left[\log \sigma\left(1-p_\omega\left(\mathbf{h}_j^{\prime}, \mathbf{h}_u^{\prime \prime}\right)\right)\right], \\
p_\omega\left(\mathbf{h}_u^{\prime}, \mathbf{h}_u^{\prime \prime}\right)=\sigma\left(\mathbf{h}_u^{\prime} \mathbf{W}\mathbf{h}_u^{\prime \prime T}\right)
\end{gathered}
\end{equation}
where $\mathbf{h}_u^{\prime}$ and $\mathbf{h}_u^{\prime \prime}$ are the coarse-grained user representation and fine-grained user representation, respectively.  $n$ is the number of negative samples. 

\textbf{SGGCF}~\cite{SGGCF} performs user node dropout and edge dropout on the user-item-group graph. Moreover, representations of the same nodes in original and augmented graphs are viewed as positive pairs. Besides, it performs cross-layer contrasting. The MI between initial node embedding and $l$-th ($l$ is an even number) layer embedding is maximized.

\textbf{Sequential Recommendation.}
\textbf{DCRec}~\cite{DCRec} constructs sequential and collaborative views based on interaction sequences. It builds an item transition graph and an item co-interaction graph as collaborative views. Additionally, DCRec adjusts the strength of contrastive regularization by disentangling user conformity and actual interest, further enhancing its performance, which is formulated as: 
\begin{equation}
\begin{gathered}
 \mathcal{L}_{con} =-\sum_{i \in s_u} \omega_{u, i} \log \frac{\exp \left(\cos \left(\mathbf{h}_i^s, \mathbf{h}_i^t\right) / \tau\right)}{\sum_{i^{\prime} \in \mathcal{\mathcal{I}}} \exp \left(\cos \left(\mathbf{h}_i^s, \mathbf{h}_{i^{\prime}}^t\right) / \tau\right)} \\
 - (1-\omega_{u, i}) \log \frac{\exp \left(\cos \left(\mathbf{h}_i^t, \mathbf{h}_i^c\right) / \tau\right)}{\sum_{i^{\prime} \in \mathcal{\mathcal{I}}} \exp \left(\cos \left(\mathbf{h}_i^t, \mathbf{h}_{i^{\prime}}\right) / \tau\right)} 
\end{gathered}
\end{equation}
where $s_u$ is the interaction sequence of user $u$. $\mathbf{h}_i ^s$, $\mathbf{h}_i ^t$, $\mathbf{h}_i ^c$ are representations of item $i$ in the sequence view, item transition graph and item co-interaction graph, respectively. $\omega_{u,i}$ is the conformity degree of $(u, i)$.

\textbf{Session-based Recommendation.}
Based on the session data, \textbf{COTREC}~\cite{COTREC} constructs two graph views, including item view (item-item graph) and session view (session-session graph). Inspired by SEPT, it utilizes co-training~\cite{co_training} to predict the positive and negative samples for each session. 
Note that the positive and negative samples are items. 
Furthermore, it maximizes the agreement between the representation of the last item in the session and representations of the predicted positive samples. 
The agreement between the representation of the last item and representations of the negative samples is minimized.

\textbf{Bunndle Recommendation.} Based on user-item interaction, user-bundle interaction, and item-bundle 
affiliation, \textbf{CrossCBR}~\cite{CrossCBR} constructs a bundle view (user-bundle graph) and an item view (user-item graph and bundle-item graph). It performs edge dropout and message dropout on these views and generates user representation and bundle representations. Moreover, MI between representations of the same user/bundle in corresponding views is maximized.

\subsubsection{Contextual-Contextual Contrasting}
For contextual-contextual contrasting, discrimination is performed on the contextual representations. It can be formulated as:
\begin{equation}
    \theta^{*}, \omega^{*}=\underset{\theta, \omega}{\arg \min } \mathcal{L}_{con}\left(p_\omega\left(\mathbf{c}_i, \mathbf{c}_j\right)\right)
\end{equation}
where $\mathbf{c}_i$ and $\mathbf{c}_j$ are contextual representations denoting data with similar contextual information.

\textbf{HIN-based Recommendation.}
\textbf{CHEST}~\cite{CHEST} extracts subgraphs from heterogeneous information networks (HIN) by considering path relevance and then applies data-based augmentations to them. Augmented subgraphs generated from the same original subgraphs are considered positive samples, while subgraphs connecting the same user to other items are considered negative samples. Furthermore, it employs generative pretext tasks, such as masked node/edge prediction, to capture local information. The model leverages curriculum learning~\cite{bengio2009curriculum} by pre-training in a progressive manner, with contrastive pretext tasks serving as the advanced course and generative pretext tasks as the elementary course.
\textbf{KGIC}~\cite{KGIC} constructs local and non-local graphs by combining user-item interactions and the knowledge graph. Moreover, \textit{intra-view} contrasting and \textit{inter-view} contrasting are performed among these graphs.

\textbf{Multi-behavior Recommendation.}
 To capture different user behavior patterns, \textbf{HMG-CR}~\cite{yangHyperMetaPathContrastive2021} performs the hyper meta-graph discrimination task.
It first constructs different hyper meta-graphs $\{\mathcal{G}_u^{(i)}\}_{i=1}^{K}$ for each user based on hyper meta-paths. In specific, the hyper meta-path is constructed based on the distance to the target behavior.
Then representations of these hyper meta-graphs are generated through different encoders, i.e., $\mathbf{c}_u^{(i)} = f_\theta^i(\mathcal{G}_u^{(i)})$.
Furthermore, it treats the representations of adjacent hyper meta-graphs ($\mathbf{c}_u^{(i-1)}, \mathbf{c}_u^{(i)}$) as negative pairs. The positive samples are generated by feeding the current hyper meta-graph into the encoder of the adjacent hyper meta-graph. For example, the positive sample of $\mathbf{c}_u^{(i)}$ is $\mathbf{c}_p = f_\theta^{i-1}(\mathcal{G}_u^{(i)})$. 

\textbf{Sequential Recommendation.} 
\textbf{GCL4SR}~\cite{GCL4SR} obtains subgraphs based on uniform sampling. Specifically, it first constructs a transition graph based on all user interaction sequences. For each sequence node, it randomly samples two different subgraphs. The subgraphs of the same sequence are positive pairs, and those of different sequences are viewed as negative pairs.

Unlike other works that perform contrasting on the graph, \textbf{MISS}~\cite{MISS} performs it on the feature vectors, which consist of categorical and sequential features.
It leverages two CNN-based models to extract multiple interests contained in each feature vector. 
Moreover, it makes representations of the same interest similar and representations of different interests dissimilar.

\subsubsection{Global-Global Contrasting}
Methods under this category discriminate the global representations, which can be represented as:
\begin{equation}
   \theta^{*}, \omega^{*}=\underset{\theta, \omega}{\arg \min } \mathcal{L}_{con}\left(p_\omega\left(\mathbf{g}_i, \mathbf{g}_j\right)\right)
\end{equation}
where $\mathbf{g}_i$ and $\mathbf{g}_j$ are the global representation. 
Moreover, global-global contrasting is typically used in sequential recommendation and session-based recommendation, where $\mathbf{g}$ represents a sequence or a session.

\textbf{Sequential Recommendation.}
We introduce the methods for sequential recommendation according to the view generation strategies. Generally, existing methods adopt \textit{data-based} augmentation and \textit{model-based} augmentation.

(i) \textit{Methods using Data-based Augmentation.}
\textbf{CL4SRec}~\cite{CL4SRec} utilizes three types of sequence-based augmentation: sequence cropping, sequence shuffling, and item masking. 
Given an interaction sequence of user $u$, 
it applies augmentation $\mathcal{T}$ that is randomly sampled from three augmentations to generate different sequence views. $\tilde{s}_u^{(1)} = \mathcal{T}(s_u)$ and $\tilde{s}_u^{(2)} = \mathcal{T}(s_u)$. 
SASRec~\cite{SASRec} is used as sequence encoder $f_\theta(\cdot)$ to generate sequence (global-level) representations $\mathbf{g}_u^{(1)} = f_\theta(\tilde{s}_u^{(1)})$ and $\mathbf{g}_u^{(2)} = f_\theta(\tilde{s}_u^{(2)})$. 
It performs the sequence self-discrimination task. Specifically, it makes the representations of augmented sequences from the same sequence (i.e., positive pairs) to be similar and those from different sequences (i.e., negative pairs) to be dissimilar. 
\begin{equation}
\mathcal{L}_{con}=- \log \frac{\exp (p_\omega(\mathbf{g}_u^{(1)}, \mathbf{g}_u^{(2)}))}{\exp (p_\omega(\mathbf{g}_u^{(1)}, \mathbf{g}_u^{(2)}))+Neg}.
\end{equation}
where $p_\omega(\cdot)$ is the cosine similarity with temperature parameter $\tau$. 
CL4SRec adopts in-batch negative sampling and the $Neg$ is defined as
\begin{equation}
    Neg = \sum_{v \in \mathcal{B}, v \neq u} \exp (p_\omega((\mathbf{g}_u^{(1)}, \mathbf{g}_v^{(1)})) + p_\omega((\mathbf{g}_u^{(1)}, \mathbf{g}_v^{(2)})))
\end{equation}
Similar to CL4SRec, \textbf{H$^2$SeqRec}~\cite{H2SeqRec} contrasts sequence representations. Moreover, the model is pre-trained with contrastive tasks.
\textbf{CoSeRec}~\cite{CoSeRec} and \textbf{ContraRec}~\cite{ContraRec} adopt the same framework and objective as CL4SRec. Moreover, CoSeRec proposes two robust augmentation strategies, i.e., \textit{item substituting} and \textit{item inserting}. In addition to the augmented sequence from the same sequence, ContraRec also treats the sequences that have the same target item as positive samples. Based on CoseRec, \textbf{TiCoseRec}~\cite{TiCoseRec} proposes five data augmentation based on time intervals to generate uniform sequences. Moreover, a uniform (un-uniform) sequence is one in which the standard deviation value of its time interval series is relatively small (large).
\textbf{IOCRec}~\cite{IOCRec} generates $K$ intention representations for each augmented sequence. It maximizes MI between representations of the same sequence with the same intent, which can be formulated as:
\begin{equation}
\mathcal{L}^{k}_{con}=-\log \frac{\exp (p_\omega(\mathbf{g}_{u,k}^{(1)},\mathbf{g}_{u,k}^{(2)}))}{\sum_{g \in \mathcal{N}} \exp (p_\omega(\mathbf{g}_{u,k}^{(1)}, \mathbf{g}))} .
\end{equation}
where $\{\mathbf{g}_{u,k}^{(1)}\}_{k=1}^K$ and $\{\mathbf{g}_{u,k}^{(2)}\}_{k=1}^K$ denote the intention representations of augmentation sequences. $p_\omega(\cdot)$ is the dot product. 
$\mathcal{N}$ is the set of negative samples, including different intention representations of the same user and all intention representations of different users.

\textbf{MCCM}~\cite{MCCM} incorporates contrastive learning into news recommendation. It augments sequences by performing item masking/substituting. Moreover, the masking/substituting probability is calculated based on the frequency of news. 
\begin{equation}
p_i=\frac{\log (\operatorname{count}(i))}{\max _{j \in \mathcal{I}} \log (\operatorname{count}(j))}(p_{\max }-p_{\text {min }})+p_{\text {min }}
\end{equation}
where $p_{\max}$ and $p_{\min}$ are the predefined boundaries of the probability. $\operatorname{count}(i)$ is the frequency of news $i$ in the dataset. $\mathcal{I}$ is the news set. It performs sequence self-discrimination similar to CL4SRec. 

\textbf{CCL}~\cite{CCL} augments sequences by leveraging a data generator based on the mask-and-fill operation. In specific, it first masks a portion of items in each sequence. Then the data generator recovers the original
sequence. The original sequence and recovered sequences from the same user are treated as positive pairs and other recovered sequences from different users are negative samples. Moreover, CCL utilizes curriculum learning to conduct contrastive learning via an easy-to-difficult process.

In addition to interaction data, \textbf{MIC}~\cite{MIC} utilizes attributes of user/items.
It constructs user/item sequences, which consist of user/item attributes and interaction records. Then, feature dropout is applied to generate different user/item sequences. Moreover, it treats k-nearest neighbors of the user/item as positive samples. It also clusters the users/items using k-means++ and treats users/items that are from different clusters as negative samples. 

\textbf{EC4SRec}~\cite{EC4SRec} further incorporates explanation methods into data-based augmentation. It obtains the importance scores of items to guide the augmentation based on explanation methods. The importance scores of items in sequence $s_u$ are calculated as
\begin{equation}
{score}(s_u)= F_e(y_u, s_u, f_\theta), 
\end{equation}
where $F_e$ is any explanation method. $f_\theta$ is the sequence encoder. $y_u$ is the prediction probability for the next item.
${score}(s_u) = [{score}(i_{{u},{1}}),\cdots,{score}(i_{u,{|s_u|}})]$ and ${score}(i_{{u},{1}})$ is the important score of item $i_{u, 1}$. 
Then, it crops/masks/shuffles the items with the lowest scores to generate positive samples. 
It also adopts supervised positive sampling to sample sequences like ContraRec and takes the sequences with higher important scores among them as the positive samples. 
To generate negative samples, it masks the items with the highest scores. The cropped items also form negative samples. In addition to making positive samples from the same users similar, it also makes negative samples from different users more similar than positive samples from all users.

(ii) \textit{Methods using Model-based Augmentation.}
\textbf{DuoRec}~\cite{DuoRec} obtains different sequence representations based on message dropout. Specifically, the sequence is fed twice with different dropout masks in the Transformer-based encoder. Then it performs sequence self-discrimination on these representations. Similar to ContraRec, DuoRec also uses supervised positive sampling, where an interaction sequence with the same target item is randomly selected as a positive sample.

Several methods also apply dropout to generate different sequence representations. Besides message dropout, \textbf{CBiT}~\cite{CBiT} also uses item masking. Specifically, it first generates $K$ different sequences by item masking. Then, these sequences are fed into bidirectional Transformers with different dropout masks to generate representations. In addition, CBiT proposes multi-pair contrastive learning. All $K$ augmented sequences are treated as positive samples. It defined the loss as
\begin{equation}
\mathcal{L}_{\text {con }}=- \sum_{x=1}^{K} \sum_{y=1}^{K} \mathbf{1}_{[x \neq y]}   \log \frac{\exp (p_\omega(\mathbf{g}_u^{(x)}, \mathbf{g}_u^{(y)}))}{\exp (p_\omega(\mathbf{g}_u^{(x)}, \mathbf{g}_u^{(y)}))+ \sum_{\mathbf{g} \in \mathcal{N}}\exp (\mathbf{g}_u^{(x)}, \mathbf{g})}
\end{equation}
where $\mathcal{N}$ is the set of negative samples, which are the sequence representations of different users.

\textbf{ContrastVAE}~\cite{ContrastVAE} incorporates contrastive learning into the Variational
AutoEncoder. Besides message dropout and data-based augmentations, it utilizes the variational dropout that introduces a learnable Gaussian dropout rate during the sampling step. The selection of negative and positive pairs is similar to CL4SRec. \textbf{CLUE}~\cite{CLUE} also adopts both message dropout and sequence-based augmentations. The main difference is that CLUE does not use negative samples. 
\textbf{FDSA\_CL}~\cite{FDSA_CL} constructs feature sequences consisting of item features. It maximizes the MI between the representations of feature sequences and interaction sequences. In addition, it applies message dropout to generate $K$ representations for each feature sequence. 

\textbf{EMKD}~\cite{EMKD} combines contrastive knowledge distillation with ensemble modeling. Specifically, it first generates $K$ sequences by item masking. Then, it uses $N$ networks that have the same structure but different initializations as an ensemble of sequence encoders. Each parallel network is a bidirectional Transformer encoder. EMKD conducts both intra- and inter-network contrasting. The intra-network contrasting is conducted by maximizing the MI between representations of the original sequence and the augmented sequence generated by the same network. The inter-network contrasting is conducted by maximizing the MI between sequence representations generated by the different networks.

\textbf{MCLRec}~\cite{MCLRec} utilizes a learnable model-based augmentation method for view generation. It employs two different MLP-based augmenters to obtain representations to perform contrastive learning. Specifically, it first generates two sequences using data-based augmentation methods such as item masking and obtains their original representations through a shared encoder $f_\theta\left(\cdot\right)$. Then, these representations are fed into the augmenters to obtain augmented representations. MCLRec consequently contrasts the four representations. Additionally, it performs a meta-learning strategy to train the augmenters.

\textbf{Session-based Recommendation.}
\textbf{DHCN}~\cite{DHCN} constructs two hypergraphs (i.e., $\mathcal{G}_h$ and $\mathcal{G}_l$) to represent intra- and inter-session information, respectively. 
In DHCN, representations of the same session (e.g., $\mathbf{g}_s^{h}$ and $\mathbf{g}_s^{l}$) is the positive pair. 
Moreover, it perturbs the representations matrix $\mathbf{H}^h$ with row- and column-wise shuffling.  
The negative samples are the perturbed representation $\tilde{\mathbf{g}}_s^{h} \in \tilde{\mathbf{H}}$ of the same session. $\tilde{\mathbf{H}}$ is the perturbed representation matrix. It maximizes MI between positive pairs and minimizes MI between negative pairs by
\begin{equation}
\mathcal{L}_{con}=-\log \sigma(p_\omega(\mathbf{g}_s^{h}, \mathbf{g}_s^{l}))-\log \sigma(1-p_\omega(\tilde{\mathbf{g}}_s^{h}, \mathbf{g}_s^{l}))
\end{equation}
where $p_\omega(\cdot)$ is the dot product and $\sigma(\cdot)$ is the sigmoid function.

\textbf{OD-Rec}~\cite{on_device} incorporates contrastive learning into knowledge distillation for session-based recommendation. Specifically, it maximizes the mutual information between the representations of the same session $s$ that learned from the teacher model and student model (i.e., $\mathbf{g}_s^{tea}$, and $\mathbf{g}_s^{stu}$). The negative pairs are the $(\mathbf{g}_s^{tea}, \mathbf{g}_{s^\prime}^{tea})$. 
\textbf{CGL}~\cite{CGL_tois22} constructs a global graph based on the similarity of sessions. In the graph, each session node is connected with their $M$ most similar sessions. It maximizes the MI between a session node and its neighbors and minimizes the MI of unconnected sessions. Moreover, the session representation is obtained by aggregating the representations of items in it. 

\textbf{Feature-based Recommendation.}
\textbf{CFM}~\cite{CFM} adopts feature dropout to augment feature vectors. The vectors generated from the same feature vector are treated as positive pairs, while those generated from different feature vectors are treated as negative pairs.
Moreover, to make the pretext task more difficult, it masks/drops the related features. 
\textbf{CL4CTR}~\cite{CL4CTR} 
performs feature-based augmentation strategies to generate two different feature vectors of each sample.
It minimizes the expected distance between representations of the same samples and does not use negative samples. 
Besides using feature representations, \textbf{CLCRec}~\cite{CLCRec} also generates item collaborative representations based on the interaction data. Therefore, it can be viewed as \textit{hybrid recommendation}~\cite{rec_survey}.
The 
feature representation and collaborative representation of the same item are treated as 
positive pairs and those of different items are treated as negative pairs. 

\begin{figure}
    \centering
    \includegraphics[width=0.4\linewidth]{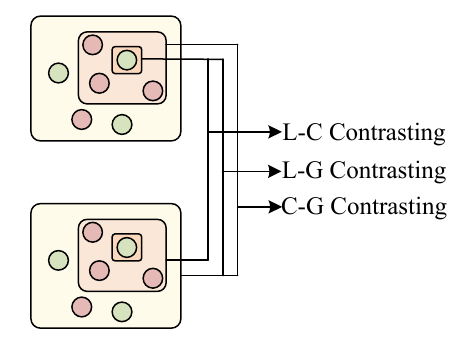}
    \caption{Illustration of cross-scale contrasting.}
    \label{fig:cross_scale_whole}
\end{figure}

\subsection{Cross-Scale Contrasting}
In the cross-scale contrasting (shown in Fig.\ref{fig:cross_scale_whole}), the contrasting is conducted across different scales.
Furthermore, according to the scale, we further divide this branch of methods into three sub-types: local-contextual (L-C) contrasting, local-global (L-G) contrasting, and contextual-global (C-G) contrasting.
\subsubsection{Local-Contextual Contrasting}
The local-contextual contrasting can be formulated as:
\begin{equation}
   \theta^{*}, \omega^{*}=\underset{\theta, \omega}{\arg \min } \mathcal{L}_{con}\left(p_\omega\left(\mathbf{h}_i, \mathbf{c}_j\right)\right)
\end{equation}
where $\mathbf{h}_i$ is the local representation and $\mathbf{c}_j$ is the contextual representation.

\textbf{Graph-based Collaborative Filtering.}
To capture contextual information, \textbf{NCL}~\cite{linImprovingGraphCollaborative2022a} proposes a prototype-contrastive objective. In specific, for each item/user, the positive sample is the prototype of the cluster it belongs to, and the negative sample is the prototype of other clusters. The prototype is the representation of the cluster center. Moreover, the prototype-contrastive objective is learned with Expectation-Maximization (EM) algorithm. NCL also performs cross-layer contrasting. For each user/item, corresponding representations output from the even-numbered layer GNN are treated as positive samples.

\textbf{Sequential Recommendation.}
\textbf{ICL}~\cite{chenIntentContrastiveLearning2022} relies on a framework similar to that of NCL~\cite{linImprovingGraphCollaborative2022a}. 
The difference is that NCL clusters the representations of users or items, while ICL clusters the representations of sequences. Specifically, in ICL, the representation of a sequence can be viewed as a local view of the cluster it belongs to. In addition, each prototype represents the user intent.
Besides, ICL also places contrasting between sequences like sequential recommendation methods in global-global contrasting.

\textbf{Cross-domain Recommendation.} \textbf{SITN}~\cite{SITN} only performs cross-domain contrasting. Like ICL, it clusters user sequence representations to represent the user interests and maximizes the MI between the representations of users and clusters. Specifically, the positive pairs are representations of a user and its corresponding cluster in different domains. The negative samples are new clusters. Additionally, SITN also maximizes the MI of user representations in different domains.

\textbf{Social Recommendation.}
In addition to using clustering algorithms, contextual information is also modeled based on human prior knowledge.
\textbf{MHCN}~\cite{MHCN} designs three types of triangle motifs based on social relations and models them with a multi-channel hypergraph encoder. Moreover, a multi-channel hypergraph is proposed to model the information. 
In each channel, MHCN hierarchically maximizes the mutual information between the user representation, the user-centered sub-hypergraph representation, and the hypergraph representation.
\textbf{SMIN}~\cite{longSocialRecommendationSelfSupervised2021a} constructs the context by generating a substructure-aware adjacent matrix based on the addition operations of different order adjacent matrices.

\textbf{Group Recommendation.}
\textbf{CubeRec}~\cite{CubeRec}  maximizes the MI between the representation of group intersection and the representations of users belonging to that intersection. Specifically, it generates a hypercube representation for each group. The group intersection representations can be viewed as contextual representations. It considers the representations of users within the intersection and the representation of the intersection as positive pairs. As such, CubeRec enhances the representations, allowing for better modeling of the common interests among different groups.

\subsubsection{Local-Global Contrasting}
This branch of methods conducts contrastive tasks between the local representation and global representation, which can be presented as
\begin{equation}
    \theta^*, \omega^* = \underset{\theta, \omega}{\arg \min } \mathcal{L}_{con}\left(p_\omega\left(\mathbf{h}_i, \mathbf{g}_j\right)\right)
\end{equation}
where $\mathbf{h}_i$ is the local representation and $\mathbf{g}_j$ is the global representation. 

\textbf{Graph-based Collaborative Filtering.} \textbf{EGLN}~\cite{yang2021egln} places the contrasting across the edge representation (i.e., the concatenation of representations of its connected nodes) and the global graph representations (i.e., the average of all edge representations). Specifically, the positive samples of $\mathbf{g}_1$ (representation of $\mathcal{G}^{(1)}$) are the edge representations in the graph $\mathcal{G}^{(1)}$ and the negative samples are the edge representations in the augmented graph $\mathcal{G}^{(2)}$. \textbf{HGCL}~\cite{cai2022hgcl} constructs node-type specific homogeneous graphs to preserve the heterogeneity. Following DGI~\cite{DGI}, it maximizes the MI between a node representation and corresponding graph presentation. Moreover, to incorporate the relationship between different node types, HGCL also designs a cross-type contrasting object. For each node type pair $(t_1, t_2)$, given the node-type specific homogeneous graph $\mathcal{G}^{(t_2)}$, the positive samples of it are the node representations in $\mathcal{G}^{(t_2)}$, and the negative samples are the node representation in the augmented graph of $\mathcal{G}^{(t_1)}$.

\textbf{Group Recommendation.} For group recommendation, there are typically two views: the user view and the group view, to perform contrastive tasks. \textbf{GroupIM}~\cite{GroupIM} maximizes the MI between the representations of group members and the representations of groups. In specific, it treats the user representation $\mathbf{h}_i$ and group representation $\mathbf{g}_j$ as positive pairs, where user $i$ belongs to group $j$. Negative samples $\mathbf{h}_{\tilde{u}}$ are sampled from non-member user representations. Moreover, GroupIM introduces a preference-biased negative user sampling distribution $\mathcal{P}_{\mathcal{N}}(\tilde{u} \mid j)$. This distribution gives a higher likelihood to non-member users who have interacted with similar items as the target group $j$. The sampling distribution is defined as:
\begin{equation}
\left. \mathcal{P}_{\mathcal{N}}(\tilde{u} \mid j) \propto \eta \mathcal{I}\left(\mathbf{x}_{\tilde{u}}^T \cdot \mathbf{x}_j>0\right\}\right)+(1-\eta) \frac{1}{|\mathcal{U}|}
\end{equation}
where $\mathbf{x}_{\tilde{u}}$ and $\mathbf{x}_{j}$ are the interacted items of user $\tilde{u}$ and group $j$, respectively. $\mathcal{I}(\cdot)$ is the indicator function. $\eta$ is the hyperparameter controlling the sampling bias, and $|\mathcal{U}|$ is the number of users.
\subsubsection{Contextual-Global Contrasting} These methods contrast the contextual representation with global representation, which can be defined as: 
\begin{equation}
    \theta^*, \omega^* = \underset{\theta, \omega}{\arg \min } \mathcal{L}_{con}\left(p_\omega\left(\mathbf{c}_i, \mathbf{g}_j\right)\right)
\end{equation}

\textbf{Graph-based Collaborative Filtering.}
For each edge $(i, j)$, \textbf{BiGI}~\cite{cao2021bigi} performs ego-net sampling to get two subgraphs centered at $i$ and $j$, respectively. Then, it adopts attention mechanism to obtain two contextual representations. The contextual representation of this edge $\mathbf{s}_{ij}$ is the concatenation of contextual representations of $i$ and $j$. Specifically, positive samples of $\mathbf{g}_1$ are the contextual representations of edges in $\mathcal{G}^{(1)}$ and negative samples are the contextual representation of edges in the augmented graph $\mathcal{G}^{(2)}$. 
\textbf{MMSSL}~\cite{MMSSL} generates multiple modal-specific representations of users. Moreover, it maximizes MI between the modality-specific representation and the overall representation of the same user. The negative samples of a modality-specific representation $\mathbf{c}_u^m$ are both the modality-specific representation $\mathbf{c}_{u^\prime}^m$ and overall representation $\mathbf{g}^m_{u^\prime}$ of different users.

\textbf{Cross-domain Recommendation.}
\textbf{C$^2$DSR}~\cite{C2DSR} obtains cross-domain sequences through merging single-domain sequences in chronological order.
Then, it generates two augmented cross-domain sequences based on item substituting. Based on the self-attention mechanism, it generates item representations in the sequences. The representations of cross-domain sequences are obtained by aggregating the representations of items in each domain. Similarly, the representations of single-domain sequences are obtained. It maximizes the MI between the single-domain sequences and the original cross-domain sequences and minimizes the MI between the single-domain sequences and the augmented cross-domain sequences.

\textbf{Sequential Recommendation.}
\textbf{SSI}~\cite{SSI} leverages contrastive learning to capture global consistency in sequential recommendation. 
Specifically, it samples a subsequence from the interaction sequence and masks the corresponding items in that sequence. It then maximizes the MI between the representation of the subsequence and that of the entire sequence. Negative samples are subsequences sampled from other sequences.
\textbf{SESRec}~\cite{SESRec} contrasts the recommendation sequence with the search sequence. It divides each sequence into a positive subsequence and a negative subsequence. Moreover, subsequences are generated based on similarity scores, which are obtained through the co-attention technique~\cite{Co-Attention}. The higher-scoring elements of the sequence are assigned to the positive subsequence while the lower-scoring elements are assigned to the negative subsequence. Anchor is generated from the original sequence. It makes anchors similar to positive sequences and different from negative sequences through the triplet loss.

Moreover, based on the belonging relationships in interaction sequences, \textbf{S$^3$-Rec}~\cite{zhouS3RecSelfSupervisedLearning2020} and \textbf{TCPSRec}~\cite{TCPSRec} conduct multiple contrastive tasks. Specifically, S$^3$-Rec devises four objectives, including sequence-item, sequence-attribute, item-attribute, and sequence-subsequence mutual information maximization.
TCPSRec performs item-sequence and item-subsequence contrasting as well as subsequence-subsequence contrasting at both coarse- and fine-grained periodicity levels. In TCPSRec, the subsequences are generated by dividing the interaction sequence when the time interval is greater than a threshold.

\begin{table*}
\caption{Comparison between different pretext tasks.}
\label{tab:pre_discussion}
\begin{tabular}{ccccccc}
\toprule
\multicolumn{1}{c|}{\multirow{2}{*}{}} & \multicolumn{3}{c|}{Same-Scale} & \multicolumn{3}{c}{Cross-Scale} \\ \cline{2-7}
\multicolumn{1}{c|}{}                  & \multicolumn{1}{c|}{L-L} & \multicolumn{1}{c|}{C-C} & \multicolumn{1}{c|}{G-G} & \multicolumn{1}{c|}{L-C}& \multicolumn{1}{c|}{L-G} & \multicolumn{1}{c}{C-G}\\ 
\midrule
Context Extraction Free& \CheckmarkBold& \XSolidBrush& \CheckmarkBold& \XSolidBrush & \CheckmarkBold & \XSolidBrush\\ 
Summary(Readout) Function Free& \CheckmarkBold& \XSolidBrush& \XSolidBrush& \XSolidBrush & \XSolidBrush & \XSolidBrush\\ 
Noise Level& \multicolumn{3}{c}{Low} & \multicolumn{3}{c}{High}\\
\bottomrule
\end{tabular}
\end{table*}

\subsection{Discussion}
Table.\ref{tab:pre_discussion} shows the comparison between different pretext tasks. Most existing methods utilize the same-scale contrasting, as it is only necessary to generate representations of the same scale. In contrast, cross-scale contrasting requires generating the corresponding representations for all the different scales. Furthermore, since existing CL-based methods usually use the shared encoder to generate representations, methods that adopt cross-scale contrasting usually require an additional module (i.e., summary function) to generate large-scale representations after generating small-scale representations. 
Take local-global contrasting in the graph-based recommendation as an example, it needs to learn the representation of each node first and then aggregate these representations to generate the graph representation using a readout function.
The contextual contrasting also tends to have high complexity, because it needs to design the corresponding strategy (i.e., context extraction) to decide which part of the data to generate the contextual representation. 

In addition, compared to same-scale contrasting, which usually aims to identify different instances, cross-scale contrasting focuses on modeling the belonging relationship between small and large scales. Considering the complexity, in the cross-scale contrasting, the negative samples are usually selected from the small-scale representations.
Moreover, cross-scale contrasting can introduce more information into the small-scale representations, but this may also introduce more noises, i.e., irrelevant information.

The objective of CL-based recommendation is to enhance recommendation performance by incorporating contrastive pretext tasks as auxiliary tasks. Therefore, the choice of pretext tasks depends on the specific recommendation tasks. For instance, in sequential recommendation where the goal is to learn sequence representations, utilizing (sub)sequence-level contrasting will perform better compared to solely relying on item-level contrasting. In recommendation tasks with inherent multiple views like KG-based recommendation, employing both inter- and intra-view contrasting can further enhance model performance~\cite{KGIC, KGCL}. This is because, under the supervision of the CL, auxiliary views such as knowledge graphs can acquire better representations.

In addition, we can address common issues by designing the sampling strategy for negative and positive samples. For example, to solve the data noise problem, one approach is to divide the data into noise-free data and noisy data and then treat them as positive and negative data for the original data, respectively~\cite{qinWorldBinaryContrastive2021}. When sampling, focusing more on tailed samples can help mitigate the bias problem~\cite{Bias_Survey}. Likely, for the cold-start problem, paying more attention to users/items that have a few interactions can be helpful.

\section{Contrastive Objective}\label{sec:obj}
As introduced in Section.~\ref{sec:unified_framework}, the contrastive objective is to maximize the mutual information (MI) between different views. Specifically, given representations $(\mathbf{h}_i, \mathbf{h}_j)$ of instances $(i,j)$, the MI between them can be represented as:
\begin{equation}
    \mathcal{M I}\left(\mathbf{h}_i, \mathbf{h}_j\right)=KL\left(P\left(\mathbf{h}_i, \mathbf{h}_j\right) \| P\left(\mathbf{h}_i\right) P\left(\mathbf{h}_j\right)\right)=\mathbb{E}_{P\left(\mathbf{h}_i, \mathbf{h}_j\right)}\left[\log \frac{P\left(\mathbf{h}_i, \mathbf{h}_j\right)}{P\left(\mathbf{h}_i\right) P\left(\mathbf{h}_j\right)}\right]
\end{equation}
where $KL(\cdot)$ is the Kullback-Leibler (KL) divergence. Contrastive learning aims to maximize the agreement between positive pairs and minimize the agreement between negative pairs. Moreover, the positive pair comes from the joint distribution $P(\mathbf{h}_i, \mathbf{h}_j)$ and the negative pair comes from the product of marginal distributions $P(\mathbf{h}_i)P(\mathbf{h}_j)$. 

Depending on whether an estimation of lower-bound of mutual information is provided, we classify the contrastive objective into bound objective and non-bound objective.

\subsection{Bound Objective}
As calculating MI directly is difficult,
lower-bounds are derived to estimate it~\cite{hjelm2018learning}, such as the Donsker-Varadhan estimator $\mathcal{M I}_{D V}$~\cite{donsker1983asymptotic, belghazi2018mutual}, the Jensen-Shannon estimator$\mathcal{M I}_{J S}$~\cite{nowozin2016f}, and the noise-contrastive estimator (InfoNCE) $\mathcal{M I}_{N C E}$~\cite{oord2018representation, gutmann2010noise}. Therefore, MI can be maximized by maximizing the lower-bound.
Moreover, of these three estimators, only
$\mathcal{M I}_{J S}$ and $\mathcal{M I}_{N C E}$ are currently used for CL-based recommendation.

\subsubsection{Jensen-Shannon Estimator.}
Compared to the DV estimator, the Jensen-Shannon (JS) estimator enables a more efficient estimation of MI. It replaces the Kullback-Leibler divergence with the Jensen-Shannon divergence. The contrastive loss based on it can be defined as 

\begin{equation}
   \mathcal{L}_{JS} = -\mathcal{M  I } _ { J S }\left(\mathbf{h}_i, \mathbf{h}_j\right)\\
  = -\mathbb{E}_{P}\left[\log \left(p_\omega\left(\mathbf{h}_i, \mathbf{h}_j\right)\right)\right] 
- \mathbb{E}_{P \times \tilde{P}}\left[\log \left(1-p_\omega\left(\mathbf{h}_i, \mathbf{h}^{\prime}_j\right)\right)\right]
\end{equation}
$\mathbf{h}_i$ and $\mathbf{h}_j$ are sampled from distribution $P$, and $\mathbf{h}^{\prime}_j$ is sampled from distribution $\tilde{P}$.
$p_\omega(\cdot)$ is the discriminator (i.e., pretext decoder), which generates the agreement score of $\mathbf{h}_i$ and $\mathbf{h}_j$. Moreover, there may be a projection head $g_\xi(\cdot)$ in $p_\omega(\cdot)$, which map representation $\mathbf{h}_i$ to $\mathbf{z}_i$. Specifically, $g_\xi(\cdot)$ can be a linear mapping, MLP, or identical mapping. The $p_\omega(\cdot)$ can be inner product $\mathbf{z}_i^T \mathbf{z}_j$, the cosine similarity $\mathbf{z}_i^T \mathbf{z}_j / (||\mathbf{z}_i|| ||\mathbf{z}_j||)$, or bi-linear transformation $\mathbf{z}_i^T \mathbf{W} \mathbf{z}_j$.

\subsubsection{InfoNCE  Estimator.}
InfoNCE is the most popular MI lower-bound adopted in CL-based methods for recommendation. The contrastive loss based on it can be formulated as
\begin{equation}
\mathcal{L}_{NCE} = -\mathcal{M I}_{N C E} \left(\mathbf{h}_i, \mathbf{h}_j\right)=-\mathbb{E}_{P}\left[p_\omega\left(\mathbf{h}_i, \mathbf{h}_j\right)\right.\left.-\mathbb{E}_{K \sim \tilde{P}^N}\left[\log \frac{1}{N} \sum_{\mathbf{h}_j^{\prime} \in K} e^{p_\omega\left(\mathbf{h}_i, \mathbf{h}_j^{\prime}\right)}\right]\right]
\end{equation}
where $K$ is the set of samples that consists of $N$ random variables identically and independently distributed from $\tilde{P}$. Generally, $p_\omega(\cdot)$ is the cosine similarity with a temperature parameter $\tau$, i.e., $p_\omega(\mathbf{z}_i, \mathbf{z}_j) = \mathbf{z}_i\mathbf{z}_j/\tau$ and $\mathbf{z}_i = \mathbf{h}_i/||\mathbf{h}_i||$. This is also known as the NT-Xent~\cite{sohn2016improved} loss.

In practice, InfoNCE is calculated on a mini-batch $\mathcal{B}$ whose size is $N+1$. Specifically, for each instance $i$ in $\mathcal{B}$, the rest $N$ instances are considered as negative samples. The loss based on InfoNCE can be
\begin{equation}
\mathcal{L}_{NCE}=-\frac{1}{N+1} \sum_{i \in \mathcal{B}}\left[\log \frac{e^{p_\omega\left(\mathbf{h}_i, \mathbf{h}_j\right)}}{\sum_{j \in \mathcal{B}} e^{p_\omega\left(\mathbf{h}_i, \mathbf{h}_j^{\prime}\right)}}\right] .
\end{equation}

\subsection{Non-Bound Objective}
In addition to the lower-bound MI estimators mentioned above, some other objectives are used to optimize contrastive learning, i.e., triplet loss and BYOL loss. However, they have not proven to be the lower-bound of MI and thus minimizing it does not guarantee to maximize mutual information.
\subsubsection{Triplet Loss.}
The triplet loss does not minimize the agreement of the negative pairs but only makes the agreement of the positive pairs greater than that of the negative pairs.
It is defined as:
\begin{equation}
\mathcal{L}_{Triplet}\left(\mathbf{h}_i, \mathbf{h}_j\right)=\mathbb{E}_{P \times \tilde{P}}\left[
\max \left\{p_\omega\left(\mathbf{h}_i, \mathbf{h}_j^{\prime}\right)-p_\omega\left(\mathbf{h}_i, \mathbf{h}_j\right)+\epsilon, 0\right\}\right]
\end{equation}
where $\epsilon$ is the margin value. $\mathbf{h}_i$ and $\mathbf{h}_j$ are sampled from distribution $P$, and $\mathbf{h}_j^{\prime}$ is sampled from $\tilde{P}$.
The discriminator $p_\omega$ can calculated the agreement by $p_\omega{(\mathbf{h}_i, \mathbf{h}_j) = \text{sigmoid}(\mathbf{h}_i, \mathbf{h}_j)}$ or $p_\omega(\mathbf{h}_i, \mathbf{h}_j) = ||(\mathbf{h}_i - \mathbf{h}_j)||$.

\subsubsection{BYOL Loss.}
This objective is proposed by BYOL~\cite{BYOL}. 
It only maximizes the agreement of positive pairs and does not use negative samples. It is defined as:

\begin{equation}
\mathcal{L}_{BYOL}\left(\mathbf{h}_i, \mathbf{h}_j\right)=\mathbb{E}_{{P} \times {P}}\left[2-2 \cdot \frac{\left[p_\psi\left(\mathbf{h}_i\right)\right]^T \mathbf{h}_j}{\left\|p_\psi\left(\mathbf{h}_i\right)\right\|\left\|\mathbf{h}_j\right\|}\right]
\end{equation}
where $\mathbf{h}_i$ and $\mathbf{h}_j$ are sampled from $P$.
$p_\psi(\cdot)$ is an online predictor. 
As it does not use negative samples to prevent collapse, other designs are needed. For example, BYOL~\cite{BYOL} utilizes momentum encoders, stop gradient, etc. 

\begin{table*}
\caption{Comparison between different contrastive objectives.}
\label{tab:ob_discussion}
\resizebox{\textwidth}{!}
{
\begin{tabular}{cccccc}
\toprule
\multicolumn{1}{c|}{\multirow{2}{*}{}} & \multicolumn{2}{c|}{Bound} & \multicolumn{2}{c}{Non-bound} \\ \cline{2-5}
\multicolumn{1}{c|}{}                  & \multicolumn{1}{c|}{Jensen-Shannon Estimator} & \multicolumn{1}{c|}{InfoNCE Estimator} & \multicolumn{1}{c|}{Triplet Loss} & \multicolumn{1}{c}{BYOL Loss}\\ 
\midrule
Lower-bound MI estimation& \CheckmarkBold& \CheckmarkBold& \XSolidBrush& \XSolidBrush\\ 
Batch-Size Independence& \CheckmarkBold& \XSolidBrush& \CheckmarkBold& \CheckmarkBold\\
Uniformity& \CheckmarkBold& \CheckmarkBold& \CheckmarkBold& \XSolidBrush\\
Low Variance& \XSolidBrush& \CheckmarkBold& N.A.& N.A.\\
\bottomrule
\end{tabular}
}
\end{table*}

\subsection{Discussion}
Table.\ref{tab:ob_discussion} shows the comparison between different contrastive objectives.
Among all the contrastive objectives, InfoNCE is the most widely used due to its good performance. Moreover, both InfoNCE and JS estimate MI based on lower-bound, and \citet{low_variance} demonstrates that InfoNCE has a lower variance of the estimated MI than JS.
However, InfoNCE requires a large number of negative samples and thus a large batch size during training. This leads to high computational and time complexity. 
In contrast, JS can achieve better performance when the batch size is small. 

Triplet loss and BYOL loss are also independent of the large batch size. However, they lack theoretical support, i.e. no theory proves that maximizing them will achieve the goal of maximizing mutual information. Moreover, triplet loss just makes the agreement of negative pairs smaller than that of positive pairs. 
Therefore, selecting informative positive/negative samples that are difficult to discriminate can lead to better performance, whereas using random or easy samples leads to poor performance. 
Additionally, triplet loss can be sensitive to the choice of margin value, hence it requires careful adjustment. 

BYOL loss is the most efficient since it does not require negative samples. 
However, BYOL loss does not contain the uniformity proposed by~\citet{uniform_alignment}, which suggests that normalized representations should be uniformly distributed over the unit hypersphere.
Hence, it easily encounters the problem of collapse. Therefore, if BYOL loss is used, additional design is usually required to prevent it.

Overall, InfoNCE is generally a good choice. If the batch size is limited, JS may be a better alternative. Triplet loss can be adopted when positive and negative pairs should not be absolutely discriminated. For the bias issue, \citet{SimGCL} demonstrate that InfoNCE can implicitly alleviate the popularity bias by making representations uniformly distributed across the unit hypersphere. This implies JS and Triplet loss can also partially mitigate the popularity. However, they use much fewer negative samples than InfoNCE, and thus may not mitigate bias as effectively. For the noise issue, re-weighting strategies~\cite{denoise_survey} can be used to assign a higger weight to reliable (noise-free) data in the loss.

In addition, there are some contrastive recommendation loss functions such as BC loss~\cite{BC-loss}, SSM loss~\cite{SSM_mm, SSM_arx}, and CCL loss~\cite{CCL-loss}. However, these works utilize contrastive learning in a supervised manner. In specific, they treat interacted user-item pairs as positive pairs and non-interacted ones as negative pairs. As we focus on contrastive self-supervised learning, we will not discuss these supervised approaches in detail.

\section{Open Issues and Future Directions}\label{sec:future}
While contrastive learning-based recommendation methods have achieved great success, there are still some open issues. In this section, we discuss these issues and outline some potential future research directions.

\subsection{View Generation}
View generation is a key component of CL-based methods. However, unlike in computer vision, where various data augmentation methods (e.g., resize, rotation, color distortion, etc.) are available, the way of generating views for CL-based recommendation is still not well explored. Specifically, most existing CL-based recommendation methods are limited to randomly removing some interactions or disrupting the order of the interaction sequence.
Moreover, these methods are often based on intuitive designs and may not be applicable to downstream recommendation tasks~\cite{SimGCL}. 
Therefore, designing more effective view generation strategies is a promising future direction.
Generally, the view generation strategies need to have the following properties: (1) Adaptability, the generated views should be adaptive to different tasks, as different tasks may use different types of data and require different information. (2) Efficiency, view generation strategies should not have high computational or time complexity. 
Moreover, dynamically updating the augmentation strategy during training is also a promising direction.

\subsection{Pretext Task}
By solving pretext tasks, the model acquires the knowledge from data for downstream tasks. Therefore, extracting useful knowledge is an important issue. For example, CGI~\cite{CGI_ib} proposes an information bottleneck-based method that enables representations to capture the minimum sufficient information for the recommended task. However, it is designed for graph-based recommendation and is difficult to apply to other recommendation tasks. Moreover, as different tasks can capture different information, learning with multiple different pretext tasks can further improve recommendation performance. It is also worthwhile to further investigate the adaptive combination of different pretext tasks for the specific recommended task.

In contrastive pretext tasks, negative samples are essential, but obtaining informative negative samples is challenging. The commonly used uniform sampling strategy, which obtains negative samples by random sampling, suffers from false negatives. Besides, easy negative samples may degrade the performance of contrastive learning as they provide little information. Therefore, effective negative sampling strategies deserve further investigation. Some works~\cite{chuang2020debiased, hard_neg_mixing} explore this problem in computer vision. However, these methods are specifically designed for image data and are difficult to apply to recommendation methods. Moreover, since current methods require a large number of negative samples, efficient negative sample strategies also need to be explored.

\subsection{Contrastive Objective}
Most CL-based recommendation methods use InfoNCE as their objective function due to its simplicity and effectiveness. Although great success has been achieved, two issues need further exploration.
First, the measurement of mutual information in InfoNCE is based on KL divergence. Therefore, it suffers from problems stemming from KL divergence (e.g., asymmetrical estimation and unstable training). Hence, better mutual information measurement is required but a few works~\cite{w_mutual} investigate this issue. 
~\citet{w_mutual} propose the Wasserstein discrepancy measurement based on the 2-Wasserstein distance to measure mutual information. In addition, it has only been applied to sequential recommendation, and its applicability to other recommendation tasks needs to be further explored.
Second, current methods use mutual information to measure the agreement. However, mutual information has several shortcomings. Besides being hard to estimate, mutual information can also lead to suboptimal representations~\cite{other_information}. Therefore, exploring alternative measures of the agreement, such as $\mathcal{V}$-information proposed by \citet{v_information}, is a promising direction.

\subsection{Miscellaneous}
\subsubsection{Meeting Real-World Recommendation}
Most existing CL-based recommendation models are trained offline. However, in real-world recommendation scenarios, such as online shopping and news recommendation, large-scale interaction data are continuously generated and user preferences are dynamic. Offline trained models may suffer from the problem of information asymmetry as they rely only on historical user interaction data to make recommendations. Hence, exploring online learning strategies in CL-based recommendation to quickly capture dynamic preference trends would be a potential direction. Moreover, conversational recommendation~\cite{sun2018conversational,zhou2022c2} is also proposed to address the information asymmetry. Specifically, the model makes recommendations based on the multi-turn interaction (e.g., dialogues) with users. By leveraging real-time user feedback, the users' current preferences can be modeled. Combining contrastive learning with conversational recommendation methods would be an interesting direction to explore.

In addition, there are various types of data noise in recommender systems, like random clicks and false interactions, which influence the effectiveness of recommendation models. Consequently, recommendation denoising has gained considerable attention. However, current CL-based recommendation methods are mainly implicit denoising. They achieve denoising by learning perturbation-invariant representations to enhance the robustness of the model. Explicit strategies to address the issue of data noise are rarely explored. One potential approach is to construct a denoised data view to perform contrastive learning. Moreover, as denoising may impair recommendation diversity, leveraging contrastive learning to balance diversity and accuracy when denoising can also be a promising direction. For instance, RGCF~\cite{RGCF} incorporates contrastive learning into recommendation denoising by maximizing the mutual information between the denoised graph and the diversity graph.
 
Apart from denoising, recommendation debiasing has also been a hot research topic in recent years. DCRec~\cite{DCRec} investigates the combination of contrastive learning and debiasing techniques. It addresses the popularity bias by disentangling user conformity and interest to adjust the contrastive regularization strength. Moreover, some existing debiased recommendation models can also be used as backbones in combination with contrastive learning to address the popularity bias issue. For example, UnKD~\cite{UnKD}, which proposes an unbiased knowledge distillation approach, can be combined with contrastive learning by incorporating contrastive learning into knowledge transfer or by treating partition groups as contrastive views. Additionally, there are various other biases, such as selection bias, and incorporating contrastive learning to mitigate these biases is also worth exploring.

Additionally, in recommender systems, data is often multi-modal, including video, text, images, etc. These modalities contain rich information and are useful for improving recommendation performance, particularly when interaction data is sparse. Therefore, deriving useful knowledge from multi-modal data through contrastive learning is a promising direction. However, only a limited number of studies~\cite{PAMD,MMSSL,SLMRec} have explored this area.

\subsubsection{Learning with Advanced Techniques}
With the rapid development of deep learning, many advanced techniques can be used to improve the performance of CL-based recommendation. One such technique is Knowledge Distillation (KD)~\cite{Knowledge_distill}, which is proposed to address the trade-off between high cost and model performance. Typically, KD first trains a large teacher model using the training set. Then a small student model is trained under supervision from the soft labels generated by the teacher model. Thus, there are naturally two views, namely the teacher view and the student view in KD. It can be easily combined with contrastive learning. Recently, several studies~\cite{EMKD, SimRec, LWC-KD} have proposed to incorporate contrastive learning into KD and achieved promising recommendation performance. However, these explorations are still at an early stage, and there is still significant research potential in combining KD with contrastive learning.
Moreover, some works~\cite{SEPT, COTREC} leverage semi-supervised learning to obtain more supervision signals. In specific, SEPT~\cite {SEPT} and COTREC~\cite{COTREC} use tri-training and co-training to obtain more informative samples, respectively.
CML~\cite{weiContrastiveMetaLearning2022} unifies contrastive learning and meta-learning by capturing meta-knowledge through contrastive learning. CCL~\cite{CCL} incorporates curriculum learning~\cite{bengio2009curriculum} into contrastive learning.

Additionally, there are still many new techniques that can be utilized for CL-based recommendation. For example, there are various choices for view generation strategies, contrastive tasks, encoders, etc. in CL-based methods. Therefore, it is a promising direction to use Automated Machine Learning (AutoML)~\cite{automl} to automatically select the appropriate method to reduce human effort.

\section{Conclusion}~\label{sec:conclusion}
In this survey, we present a comprehensive and systematic review of recent works in contrastive self-supervised learning-based recommendation. We first propose a unified framework and then introduce a taxonomy based on its key components, which include view generation strategy, pretext task, and contrastive objective. For each component, we provide detailed descriptions and discussions to guide the choice of the appropriate method. Finally, we discuss open issues and promising research directions for contrastive self-supervised learning-based recommendation in the future. We hope that this survey can provide both junior and experienced researchers with a comprehensive understanding of contrastive self-supervised learning-based recommendation and inspire future research in this area.

\begin{acks}
This research is supported in part by National Science Foundation of China (No. 62072304), Shanghai Municipal Science and Technology Commission (No. 21511104700), the Shanghai East Talents Program, the Oceanic Interdisciplinary Program of Shanghai Jiao Tong University (No. SL2020MS032), and Zhejiang Aoxin Co. Ltd.
\end{acks}

\bibliographystyle{ACM-Reference-Format}
\bibliography{sample-base}

\end{document}